\documentclass{article} 
\usepackage{graphicx}
\usepackage{color}
\newcommand{\EPPlab}{\pkg{EPP-lab}}
\newcommand{\proglang}{}
\newcommand{\pkg}[1]{\texttt{#1}}
\newcommand{\code}[1]{\texttt{#1}}
\usepackage{amsmath}
\usepackage{url}
\usepackage{Sweave}
\usepackage{natbib}
\author{Daniel Fischer \\
   Natural Resources Institute Finland(Luke)\\
   Green Technology\\
   31600 Jokioinen, Finland\\
   and\\
   School of Health Sciences\\
   University of Tampere\\
   33016 University of Tampere, Finland\\
   E-mail: \url{daniel.fischer@luke.fi}\\
  \and
   Alain Berro  \\
   Institut de Recherche en Informatique de Toulouse\\
   University of Toulouse  Capitole\\
   21 all\'ee de Brienne, 31000 Toulouse, France\\
   E-mail: \url{alain.berro@irit.fr}\\
   \and
   Klaus Nordhausen \\
   Department of Mathematics and Statistics\\
   University of Turku\\
   20014 University of Turku, Finland\\
   and\\
   School of Health Sciences\\
   University of Tampere\\
   33016 University of Tampere, Finland\\
   E-mail: \url{klaus.nordhausen@utu.fi}\\
   \and
   Anne Ruiz-Gazen \\
   Toulouse School of Economics\\
   University of  Toulouse  Capitole\\
   21 all\'ee de Brienne, 31000 Toulouse, France\\
   E-mail: \url{anne.ruiz-gazen@tse-fr.eu}\\
 }

\title{REPPlab: An R package for detecting clusters and outliers using exploratory projection pursuit}

\begin{document}

\maketitle

\section*{Abstract}
The \proglang{R}-package \pkg{REPPlab} is designed to explore multivariate data sets using one-dimensional unsupervised
projection pursuit. It is useful in practice as a preprocessing step to find clusters or as an outlier detection tool for multivariate numerical data.
Except from the package \pkg{tourr} that implements smooth sequences of projection matrices and \pkg{rggobi} that provides an interface
to a dynamic graphics package called GGobi, there is no implementation of exploratory projection pursuit tools available in R especially
in the context of outlier detection. \pkg{REPPlab} is an \proglang{R} interface for the \proglang{Java} program \EPPlab{} that implements four
projection indices and three biologically inspired optimization algorithms.
The implemented indices are either adapted to cluster or to outlier detection and the optimization algorithms have at most one parameter to tune.
Following the original software \EPPlab{}, the exploration strategy in \pkg{REPPlab} is divided
into two steps. Many potentially interesting projections are calculated at the first step and examined at the second step. 
For this second step, different tools for plotting and combining the results are proposed
with specific tools for outlier detection. Compared to \EPPlab{}, some of these tools are new and their performance is illustrated through some simulations
and using some real data sets in a clustering context. The functionalities of the package are also illustrated for outlier detection on a new data set that is provided
with the package.\\

\textbf{Keywords}: genetic algorithms, Java, kurtosis, particle swarm optimization, projection index, Tribes, projection matrix, unsupervised data analysis

\section[Introduction]{Introduction}

Exploratory Projection Pursuit (EPP) aims at finding potentially existing structures, typically clusters or outliers, in multivariate data sets by optimizing
some index that reflects the interestingness of low dimensional linear projections. Principal Component Analysis (PCA) can be seen for example as a projection
pursuit approach when using a scale measure as projection index.  However, the main idea of EPP is to go beyond traditional PCA which only focuses on second
order moments and to consider other projection indices usually measuring non-gaussianity. The founding papers of EPP \cite{friedman1974,huber1985} date back
to the seventies and eighties and proposed already several projection indices together with different strategies about how to apply them for data exploration.

Recently, there is a renewed interest for EPP in several fields like for example hyperspectral imagery \cite{malpica2008}, chemistry \cite{HouWentzell2011}
and genetics \cite{espezua2014}. In these fields, the number of variables may be high with a limited number of samples and projection pursuit is suitable
since it avoids the curse of dimensionality by projecting the data onto low dimensional subspaces. Nevertheless, as noticed by several authors, there exists
almost no implementation of static EPP tools in statistical software. The \proglang{R} packages \pkg{tourr} \cite{tourr} and \pkg{tourrGui} \cite{huang2012},
which contains a graphical user interface for \pkg{tourr}, are dedicated to tour animations while the \pkg{rrggobi} package \cite{cook2007} provides a command line interface to the
interactive and dynamic graphical package GGobi.
The package \pkg{pcaPP} \cite{pcaPP} is concerned with robust
Principal Component Analysis.

The standalone \proglang{Java} program \EPPlab{} \cite{EPPlab} is a program dedicated to EPP, described in \cite{larabi2016} and freely available on GitHub.
\EPPlab{} provides the possibility to run several
times an optimization algorithm for a chosen one-dimensional projection index and to analyze the resulting projections in detail. Different projection
indices and several biologically inspired algorithms are implemented in \EPPlab{}. The package \pkg{REPPlab} \cite{REPPlab} is an interface that gives
\proglang{R} users access to all implemented projection indices and optimization algorithms of \EPPlab{}. And in addition it provides several functions
to visualize and analyze the results from \EPPlab{} to permit a thorough exploratory analysis of the obtained projections using \proglang{R} functionalities.

In general, EPP has two essential ingredients: a projection index and an optimization algorithm. Concerning the projection index,
it is widely accepted in the EPP literature that it should measure the non-gaussianity of a projection, the gaussian distribution
being considered as the most uninteresting. There exist several families of indices as detailed in \cite{caussinus2009}, \cite{rodriguez2010}
and \cite{koch2013analysis} that are aimed at revealing different non-gaussian structures. The four indices implemented in \EPPlab{} are the Friedman-Tukey,
the Friedman, the discriminant and the kurtosis (either maximized or minimized) indices.  They belong to different families and, except for the so-called
discriminant index which is a new proposal, they are well known indices that have been studied in detail in the statistical literature.

As for the optimization algorithms, many proposals have been made for EPP
(see \cite{berro2010} for some references and also \cite{tu2003}, \cite{gou2010}, \cite{espezua2014}).
Following \cite{berro2010} and \cite{larabi2010}, \EPPlab{} implements genetic and particle swarm optimization (PSO)
algorithms that are biologically inspired. Such algorithms do no rely on any smoothness assumption of the function to optimize. Moreover the Tribe algorithm,
which is a particular PSO algorithm, is especially suited to find local optima which are of interest given the exploratory strategy we propose to follow and
describe now.

The exploration philosophy in \EPPlab{} is not one of the traditional strategies in EPP. Usually, either a dynamical approach as in \pkg{tourr} or \pkg{rggobi},
or a global
optimization method together with a structure removal \cite{friedman1987} is used. A dynamic approach is clearly of interest but it may be considered as
tedious for the data analyst. The usual alternative strategy consists in iterating a two step procedure: look for the best projection direction, namely the
one associated with the global optimum of the projection index, and remove the structure found from the data set. However, such a strategy may have some drawbacks.
For example finding a global optimum for a projection index is usually not a trivial task. This strategy disregards also all the  possible different local
optima found during this optimization process and may be time consuming. Moreover, as stated in \cite{friedman1987}, the structure removal based on orthogonal projections
may miss some interesting projections and other proposals are time consuming. The strategy
in \EPPlab{} differs from the previous one in the sense that all the local optima are saved for further analysis \cite{ruiz2010}. Such
a strategy has also been recently discussed in the discussion \cite{Villa2015,Wickham2015a} of the paper \cite{Wickham2015}.  The use of genetic and
particle swarm optimization methods helps in exploring efficiently the space of one-dimensional projections in this context.

Given the many potentially interesting projections obtained by using several projection indices and optimization algorithms, it is necessary to summarize the
information because several directions may contain the same information as already stated in \cite{friedman1987contribution}. One way to do this is by combining the different projection directions as proposed
in \cite{LiskiNordhausenOjaRuizGazen:2015}. This method basically combines several projection matrices with possibly different ranks to obtain an average
projection matrix which also automatically chooses the rank for the average projection. Using this approach for EPP leads to summarized the similar projections
in only one direction while different directions, that may correspond to some other local minima from the same or other projection indices or optimization algorithms,
are combined in separate directions. The method is illustrated in the present paper through simulations and on a real data set.

The package \pkg{REPPlab} can be used as a preliminary step before clustering. First, it may help the data analyst in checking whether the data actually contains
some clusters and thus helps in validating the use of some clustering method. EPP can also give guidelines for the choice of the number of clusters. Moreover,
clustering may be performed on projection directions rather than on the original variables. This idea will be detailed in the simulations section. Another
interesting feature of \pkg{REPPlab} is its ability to detect multivariate outliers. Observations that differ from the main bulk of the data are likely to be
revealed by some projections associated with local maxima of the kurtosis for instance. And the package incorporates some specific functions to tag
the observations that are far from the mean, or other location estimates, in terms of number of standard deviations, or other scale estimates, on the selected projection directions.

The outline of the paper is the following.  In the second section, the projection indices and the optimization algorithms implemented in \EPPlab{} are described
together with the proposed combining methodology. Then the main features of the package are given in the third section. The advantage of combining several methods
and local minima is illustrated in the fourth section using a simulation study in a clustering context. The data exploratory process using \pkg{REPPlab} is
detailed for clusters and for outlier detection in Section~\ref{Examples} using three data sets. The last section concludes the paper.

\section{Exploratory projection pursuit}

The implementation of \EPPlab{} follows the work by \cite{berro2010} and \cite{larabi2010} who showed the value of considering several projection indices and several
optimization algorithms in order to get the most of exploratory projection pursuit. It is described in \cite{larabi2016}.

\subsection{Projection indices}

When implementing \EPPlab{}, the choice was made to consider three well-known indices from different families together with a new proposal called discriminant index.
The Friedman-Tukey, the Friedman and the kurtosis indices have been widely studied in the literature but, up to our knowledge, they are not implemented in \proglang{R}.
We recall the indices definition below but more details can be found in \cite{berro2010}. We denote by $X = (x_1,\ldots,x_n)^\top$ the data set consisting of $n$ $p$-variate
observations, $u$ the $p$-dimensional unit projection vector and $I$, indexed by some initials, the projection indices.

The {\bf Friedman-Tukey index} is of the form$I_{FT}(u) =s(u)d(u)$ where $s$ depends only on the global variance structure while $d$ captures the local density of the
data \cite{friedman1974}. The term $s$ can be avoided for standardized data (see Section~\ref{standard}). The objective is to maximize this index. Note that it is minimized when the data follow a parabolic distribution
which is not far from a Gaussian distribution and thus, in some sense, it measures the departure of the projected data from normality. However, it is is known to be very sensitive
to outliers.

Using a kernel density estimate for $d$ \cite{jones1987}, this index can be written as:

\begin{equation*}
\hat I_{FT}(u) = \frac{1}{n^2h}\sum^n_{i=1}\sum^n_{j=1}K\left(\frac{u^\top (x_i-x_j)}{h}\right),
\end{equation*}
where $h$ denotes the bandwidth.
Following \cite{klinke2012} for the  kernel choice and \cite{silverman1986} for the bandwidth choice, we use the uniform kernel

\begin{equation}
K(x) = \frac{35}{32} (1-x^2)^3  \space{    }  \mathbf{1}_{\{ \arrowvert x\arrowvert \leq 1\}},
\end{equation}
where $\mathbf{1}_A(x) $ equals 1 if $x$ belongs to A and 0 otherwise and
$h = 3.12n^{-\frac{1}{6}}$.\\

The {\bf Friedman index} is an approximation of a weighted $L^2$ measure of departure of the projected data distribution from the Gaussian distribution. The objective is to
maximize this index and the approximation is based on density expansions using Legendre orthogonal polynomials. It was introduced by \cite{friedman1987} with the idea of
up-weighting distances in the center of the distribution rather than in the tails in order to detect clusters and avoid the influence of outliers. Actually, it is known that
the Friedman's index performs better than the Friedman-Tukey's index when the objective is to separate clusters. The index is an approximation of

\begin{equation}
\int_{R} (f(x)-\phi(x))^2g(x)dx,
\end{equation}
where $f$ is the density of the projected data, $\phi$ is the univariate standard normal density and $g$ is a weight function \cite{cook1993}.
This weighted $L^2$ distance between $f$ and $\phi$ is approximated  after a  variable change by the sum of the first $m$ terms of a Legendre polynomial expansion. Its expression
is:
\begin{equation}
I_{F,m}(u) = \sum^m_{j=1}\frac{2j+1}{2}\left[\frac{1}{n}\sum^n_{i=1}L_j\{2\Phi(u^\top x_i)-1\}\right]^2,
\end{equation}  \\
where $\Phi$ is the cumulative distribution function of the standardized gaussian distribution and $L_j$ is the Legendre polynomial of degree $j$, for $j=1,\ldots,m$.
\cite{friedman1987} suggested to choose $2\le m \le 6$ and  \cite{sun1993} showed that $m$ should be at least $3$. In our implementation, we use $m = 3$.

The {\bf kurtosis index} is the fourth moment of the projected standardized data and is simply defined by:
\begin{equation}
I_K(u) = \sum_{i=1}^{n} (u^\top x_i)^4.
\end{equation}
 As detailed in \cite{pena2001} for a mixture of two distributions, maximizing the kurtosis coefficient of the projected data implies detecting outliers in the projections,
 whereas minimizing the kurtosis coefficient implies maximizing the bimodality of the projections.
Thus, because local maxima and local minima may be both of interest, we consider minimization and maximization of $I_K$.

The {\bf discriminant index} was introduced in \cite{berro2010}. For standardized data it is defined as
\begin{equation}
I_D(u)= \frac{\sum_{i=1}^{N-1} \sum_{j=i+1}^{N} w(u^\top (x_i-x_j)) (u^\top (x_i-x_j))^2}{\sum_{i=1}^{N-1} \sum_{j=i+1}^{N} w(u^\top(x_i-x_j))},
\end{equation}
where $w(\cdot)$ is a decreasing and positive weight function.
Following \cite{caussinus2009}, who developed similar ideas in a different context, we use $w(x)=\exp(-x)$ in our implementation.
In the presence of clusters, this index can be seen as a measure of the within variance of the projected data but with unknown group labels.
The idea is to mimic discriminant analysis in an unsupervised context and minimize the within variance in order to detect the presence of potential clusters.
Note that potential interesting projections are associated with minima of $I_D$.

Because the Friedman-Tukey and the discriminant indices involve double sums,  their computations are time-consuming while the Friedman and the kurtosis indices are
relatively fast to compute (see \cite{larabi2016} and Section 5 below for details).

\subsection{Optimization algorithms}

The implementation of \EPPlab{} for searching local optima  is based on biologically inspired optimization algorithms. Such algorithms are known to be efficient in finding
local optima and in exploring the whole search space. Moreover, no regularity assumption is needed on the function to optimize. However, such algorithms have been sparsely used in the literature on EPP (see  \cite{tu2003}, \cite{gou2010}, \cite{espezua2014} and more references in \cite{berro2010}). A genetic algorithm and two Particle Swarm
Optimization (PSO) algorithms are available in \EPPlab{}. The genetic algorithm together with the first PSO algorithm are described in \cite{berro2010} while the second
PSO algorithm, called Tribes, is described in \cite{larabi2010}.

{\bf Genetic algorithms} are stochastic search heuristics inspired by genetics \cite{goldbert1989}. For a given optimization problem, a population of candidate solutions
is created and evolves iteratively by selection, recombination and mutation. Concerning the genetic algorithm implemented in \EPPlab{}, the size of the population has to
be fixed by the user. For this given size, the initial population is randomly generated. It is made of individuals that correspond to standardized projection vectors and
cover the search space. The function to optimize is associated with each individual and is the projection index. The evolution of the population consists of a selection
step, based on a tournament with three participants, a  recombination step, with a two-point crossover and a probability equal to 0.65, and a mutation step which permits
a random exploration of the local neighborhood of existing solutions, with a probability equal to 0.05.


Two {\bf Particle Swarm Optimization}  algorithms are implemented in \EPPlab{} as alternatives to the genetic algorithm. A PSO algorithm  \cite{eberhart1995} is a
population-based search algorithm which simulates the social behavior of birds within a flock. It differs from other evolutionary methods such as the genetic algorithms
by using some notion of cooperation between individuals of the population, called particles.

For the first implemented PSO algorithm, the size of the population has to be given by the user. Then a population of random standardized projection vectors that
correspond to the particles is generated and the search for optima is obtained by updating the position of the particles. During this iterative process, each particle
moves according to a velocity vector which is updated taking into account not only the memory of the particle but also the memory of the neighboring particles.  All
the formula and the choice of the parameters are given in \cite{berro2010}. In particular, note that a ``cosine'' neighborhood adapted to our statistical context is
used when taking into account the neighboring particles.


The {\bf Tribes} algorithm \cite{cooren2009} is a particular PSO algorithm whose main advantage is that the user does not need to precise the size of the population of
particles. The algorithm adapts itself to the complexity of the data set. In what follows, it is our preferred algorithm.

\subsection{Combining Projection directions}

As explained in the introduction, the strategy followed by EPPlab is based on the finding of several local optima by using many starting points in the optimization algorithms.
Because many of the projection directions are likely to be redundant, it is interesting to propose some methodology to analyze and summarize the set of projection directions.
Following \cite{LiskiNordhausenOjaRuizGazen:2015}, we propose to combine the different one-dimensional directions by averaging the associated orthogonal projection matrices
and obtain a few number of projection directions that will take into account the set of all projection directions found. Roughly speaking, the idea is that if many directions
are similar in the sense that the cosines between them is close to one, all these directions will be summarized in one direction in the combined projection matrix. While if
some directions differ they will constitute different directions in the combined projection matrix.

\section[REPPlab]{The R-package \pkg{REPPlab}}

The package \pkg{REPPlab} is freely available from the Comprehensive \proglang{R} Archive Network (CRAN) at \url{http://CRAN.R-project.org/package=REPPlab} and is published
under the GNU General Public Licence (GPL) 2.0 or higher licence. For the package to work it is necessary to have besides \proglang{R} also the Java Development Kit
\proglang{JDK}, version 1.4 or higher as well as the \proglang{R} packages \pkg{rJava} \cite{rJava098}, \pkg{lattice} \cite{lattice} and \pkg{LDRTools} \cite{LDRTools} installed.

\begin{figure}
\center
\includegraphics[width=0.8\textwidth]{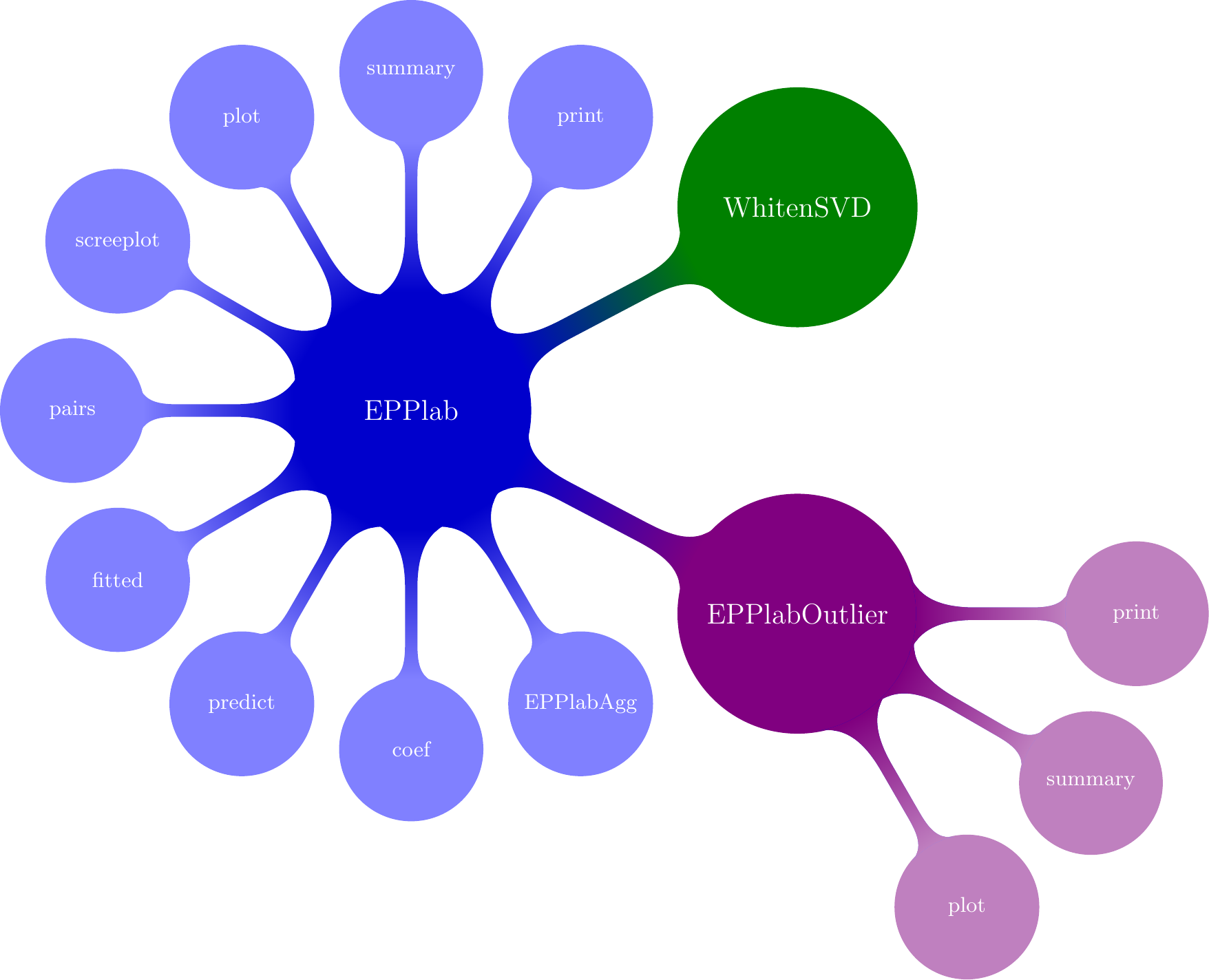}\\
\caption{Schematic overview over the functions in the \pkg{REPPlab} package.} \label{Schema}
\end{figure}

A schematic overview of the functions of \pkg{REPPlab} and their corresponding methods is provided in Figure~\ref{Schema}. The package consists mainly of the three functions
\code{EPPlab}, \code{EPPlabOutlier} and \code{WhitenSVD} where \code{EPPlab} is the main function of the package being the \proglang{R} interface to the actual \proglang{Java}
program. The most important arguments of \code{EPPlab} are the data matrix \code{x}, the desired projection index \code{PPindex} which should be computed using the algorithm
\code{PPalg} and how often the index should be computed (\code{n.simu}). The default projection index is \code{KurtosisMax}. The names of the three possible algorithms are
\code{GA} for the genetic algorithm and \code{PSO} and \code{Tribe} for the two particle swarm optimization algorithms. When specifying these arguments it is sufficient to
supply the shortest unique string.

\subsection[Standardization]{Preliminary standardization}\label{standard}

The function \code{EPPlab} always centers and scales the data, by subtracting the column means and dividing by the columns standard deviation, and the argument \code{sphere}
controls whether or not the data should also be uncorrelated. The default is \code{FALSE}, but if set to \code{TRUE} the function \code{WhitenSVD} is used to whiten the data.
\code{WhitenSVD} uses a singular value decomposition to achieve this and estimates the rank of the data as the number of eigenvalues larger than 1E-06. Therefore whitening
can also be done if there are fewer observations than dimensions.

\subsection[Input]{Input parameters for the optimization algorithms}\label{input}

To fine tune the algorithms it is possible to specify the number of particles (\code{particles}) for the standard particle swarm optimization algorithm and the number of
individuals (\code{individuals}) for the genetic algorithm. Concerning the stopping criterion, it can be modified either by specifying the maximum number of iterations
(\code{maxiter}) or by specifying the convergence criteria \verb"step_iter" and \code{eps}. The algorithms stop as soon as one of the two following conditions holds: the
maximum number of iterations is reached or the relative difference between the index value of the present iteration $i$ and the value of iteration $i$-\verb"step_iter" is
less than \code{eps}. In the last situation, the algorithm is said to converge and the function \code{EPPlab} will return the number of iterations needed to attain convergence.
If the convergence is not reached but the maximum number of iterations is attained, the function will return a warning. The default values are 10 for
\verb"step_iter" and \code{1E-06} for \code{eps}.

\subsection[Output]{Description of the outputs of the \code{EPPlab} function}\label{output}

The function \code{EPPlab} returns an object of class \code{epplab} which contains the estimated \code{n.simu} directions and their criterion value as well as other useful
information. Note that the found directions are ordered according to their criterion (index) value. For convenient working with an object of class \code{epplab}, the methods
\code{print}, \code{summary}, \code{plot}, \code{pairs}, \code{screeplot}, \code{fitted}, \code{predict} and \code{coef} are provided.

Since in usual applications of the function \code{EPPlab} the number of times the algorithms are called (\code{n.simu}) will be large it will often not be meaningful to
look at all results at once. For example \code{print} will present only information about the direction with the largest index criterion whereas all the other functions
have usually an argument \code{which} to decide which directions should be investigated. By default often only the first 10 directions will be presented.
A natural first option to explore the results is to investigate them graphically. The function \code{screeplot} can be used to compare the objective criterion values from
the different runs. The function plots the run number against its criterion value and this should give an impression which directions might actually be the same. The
\code{plot} function offers three types of plots controlled by the \code{type} argument. For \code{type = "angles"} the angles between the run with the largest objective
criterion and all other runs are  plotted. This again should give an idea which directions might actually differ and which not. The default plotting type is however
\code{type = "density"} which plots for the chosen directions marginal kernel density estimates while \code{type = "hist"} will give the corresponding histograms.
Both options indicate if the corresponding directions yield anything interesting like clusters or outliers. Similarly the function \code{pairs} produces a scatter plot matrix of the
directions included in \code{which} in order to evaluate correlations between directions or possible multidimensional structures. For further analysis of the desired
directions, the projected data can be easily extracted using the \code{fitted} function, for new observations the function \code{predict} computes the corresponding
projections. The projecting directions can be obtained using the \code{coef} function.

\subsection[Outlier]{Description of the \code{EPPlabOutlier} function}\label{outlier}

If the purpose of the analysis is outlier detection the function \code{EPPlabOutlier} is the best way to extract the relevant information from an \code{epplab} object.
In the  \code{EPPlabOutlier} function the user can specify which location and scale measure are to be used and what is the factor \code{k} that classifies an observation as outlier
based on the location and scale used. The \code{location} and \code{scale} arguments take as input, functions which return the corresponding quantities for a vector. The defaults are \code{mean}
and \code{sd} but, for example, also \code{median} and \code{mad} might be used. The \code{EPPlabOutlier} creates an object of class \code{epplabOutlier} for which \code{print},
\code{summary} and \code{plot} functions are available. The output of these three functions is most meaningful if the data given to \code{EPPlab} before calling \code{epplabOutlier}
has row labels for the observations. Just printing the object of class \code{epplabOutlier} returns a binary matrix where 1 in its $ij$-th element indicates that the observation $i$
is considered an outlier in direction $j$. A visual presentation of this binary matrix is obtained using \code{plot} where the user can choose the colors for outliers and non-outliers
and whether only those rows should be plotted which are considered at least in one direction as an outlier. A maybe more informative overview is provided using \code{summary} which
informs about the total number of outliers found and in how many directions each of the identified outliers are
considered atypical.

\subsection[aggregate]{Description of the \code{EPPlabAgg} function}\label{aggreg}


This function is designed to combine and summarize the different projection directions. The input for the function is either an \code{epplab} object or a list of such objects
when for example results from different calls with different indices or algorithms should be combined. The combining idea is quite simple. Denote as $u_i$, $i =1,\ldots, N$
the $N$ unit projection vectors which should be combined and as $P_i$,  $i =1,\ldots, N$ the corresponding projection matrices which all have rank 1. This function basically performs an
eigenvalue - eigenvector decomposition of
\[
P^* = \frac{1}{N} \sum_{i=1}^N P_i
\]
obtaining the $p$ eigenvalues $\lambda_1 \geq\ldots \geq \lambda_p$ and corresponding eigenvectors.
The idea is then to keep those $k$ eigenvectors which explain ``most'' of $P^*$. For a detailed description and the theoretic foundation see
\cite{LiskiNordhausenOjaRuizGazen:2015} who also suggest methods to automatically decide upon the value $k$. The
function \code{EPPlabAgg} offers then three options to decide upon $k$ using the \code{method} argument. Two automatic ones denoted
\code{"inverse"} and \code{"sq.inverse"} follow \cite{LiskiNordhausenOjaRuizGazen:2015} or as alternative the method \code{"cumulative"}, which
 chooses the minimal $k$ such that $\sum_{i=1}^k \lambda_i / \sum_{i=1}^N \lambda_i \geq c$ where $c$ is the percentage given by the argument \code{percentage}.


\section[Simulations]{Simulations}\label{Simulations}

EPP is often used as preprocessing step to find clusters in
data. In the following simulation study we will compare different indices
and combination methods implemented in \pkg{REPPlab} for that
purpose. For the comparison we consider two different settings
each having three underlying clusters. In the first setting the
cluster sizes are balanced with $n_1=n_2=n_3=100$  whereas in
the second setting the sizes are unbalanced with $n_1=200$,
$n_2=80$, $n_3=20$. For the cluster design we follow
\cite{HouWentzell2011} and have three 10-variate normal
populations with $N_{10}(\mu_i,\Sigma)$, where $\mu_1=(-1,
-0.58, 0, \dots, 0)^\top$, $\mu_2=(1, -0.58, 0, \dots,
0)^\top$, $\mu_3=(0, 1.15, 0, \dots, 0)^\top$ and
$\Sigma=\mbox{diag}(0.1, 0.2, 1, \dots,1)$. In order to make the design not simply visible,
we then rotate the observations with a
random orthogonal matrix. Hence when performing dimension
reduction for clustering, k=2 directions should be sufficient
to find the three clusters. The success of the different preprocessing
methods will be evaluated using the adjusted Rand index
\cite{Rand1971} when k-means clustering is used when searching
for the three clusters. The index has the range 0 to
1 and the larger the better.

For both settings we simulated 1000 times  all indices in the package \code{epplab} with the
settings \code{PPalg = "Tribe"}, \code{n.simu = 100},
\code{maxiter = 200} and \code{sphere = TRUE}.

\begin{figure} \center
\includegraphics[width=0.8\textwidth]{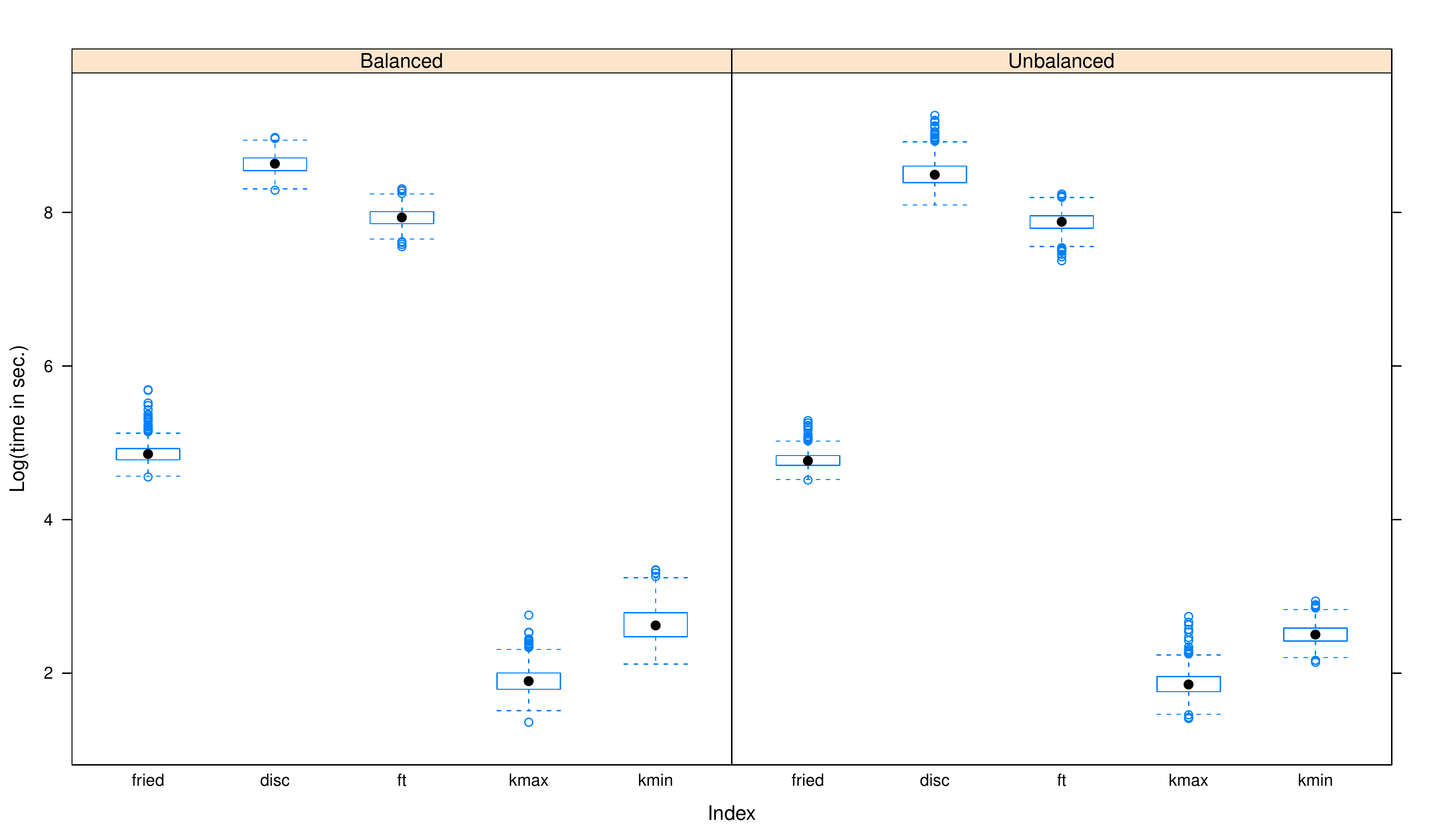}\\
\caption{Computation times for the 100 projection directions
based on 1000 repetitions in both settings.} \label{SimuTimes}
\end{figure}

To get an idea of the computational complexity the computation
times for each setting and all the indices are shown in
Figure~\ref{SimuTimes}. As can be seen here, the setting does not matter but there are
considerably differences between the indices. Both kurtosis
indices (kmin and kmax) are the fastest to compute whereas the
discriminant index (disc) and the Friedman Tukey index (ft) are
quite slow to compute. The Friedman index (fried) is intermediate in terms of computation time.

Therefore, when evaluating the performance of combination
methods we either combine all indices or only the fast
indices designed for clustering (kmin and fried) and the two combinations will be denoted as `all'
and `fast' respectively in what follows.

When combining results we used the in-build methods as provided
by \code{EPPlabAgg}. We used options \code{"inverse"},
\code{"sq.inverse"} and \code{"cumulative"} with percentages 85\%
and 95\%, denoted as `cum85' and `cum95'. In the following we
will show however only the results of inverse and cum85 as
using \code{sq.inverse} was a little worse than inverse and cum95
was clearly the worst from all of them.

\begin{figure} \center
\includegraphics[width=1.0\textwidth]{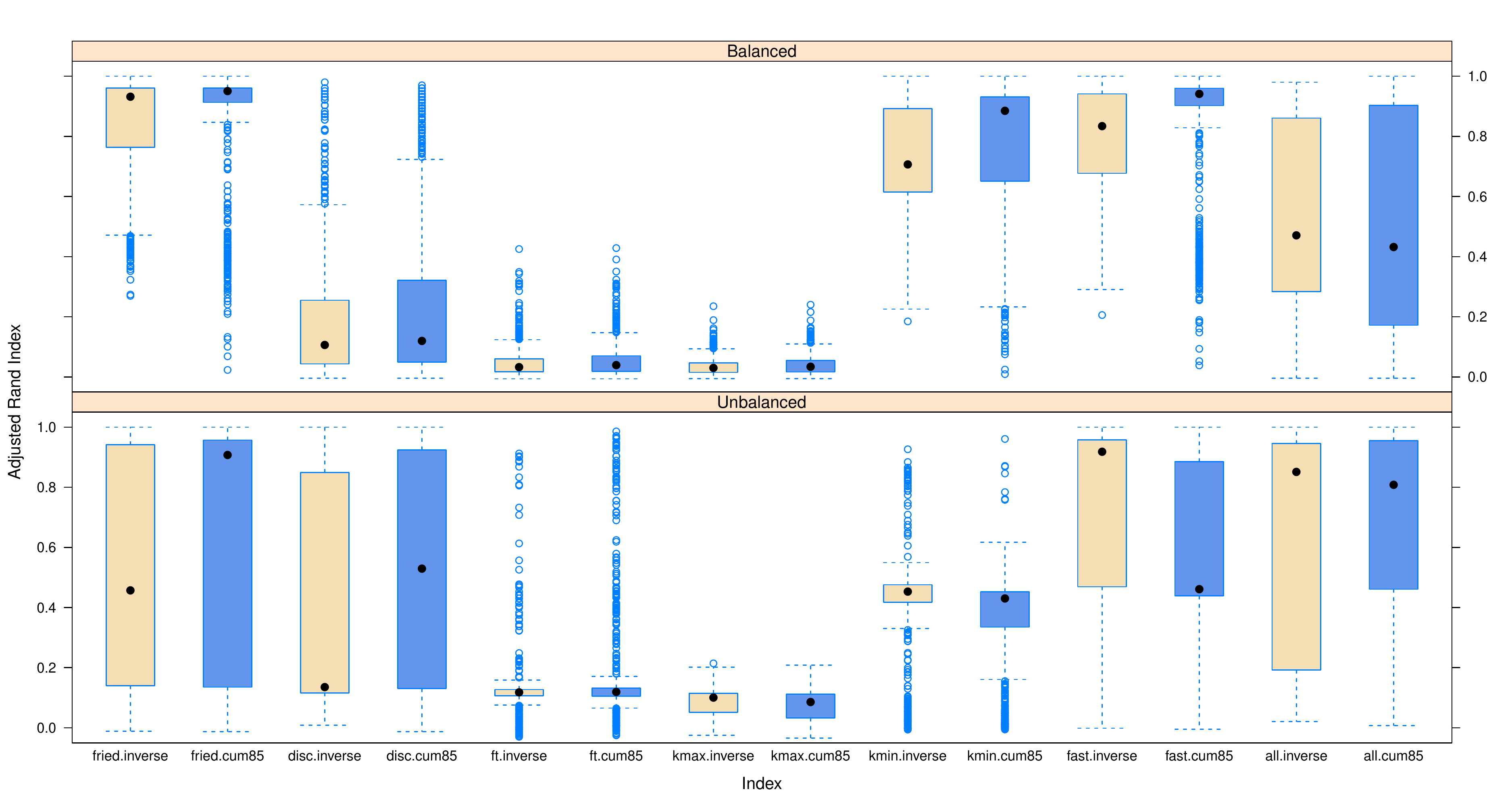}\\
\caption{Adjusted Rand index for the different combinations of
the found projections based on 1000 repetitions in both
settings (balanced at the top and unbalanced at the bottom).} \label{SimuRandTribe} \end{figure}

Figure~\ref{SimuRandTribe} shows the performance for both cases.
Obviously, maximizing kurtosis is not a good idea as a
preparation for clustering since the index is designed for finding
outliers. The Friedman-Tukey index is also not very good and the
discriminant index differs a lot between the two settings.

However when combining all  indices the performance
is in general good, but better with only the fast indices. Note that, when
combining all methods, the performance does not fail although
bad performing indices are included. When comparing the two
aggregation methods it seems that for the balanced case cum85
is better while for the unbalanced case the inverse method is better.

\begin{figure} \center
\includegraphics[width=1.0\textwidth]{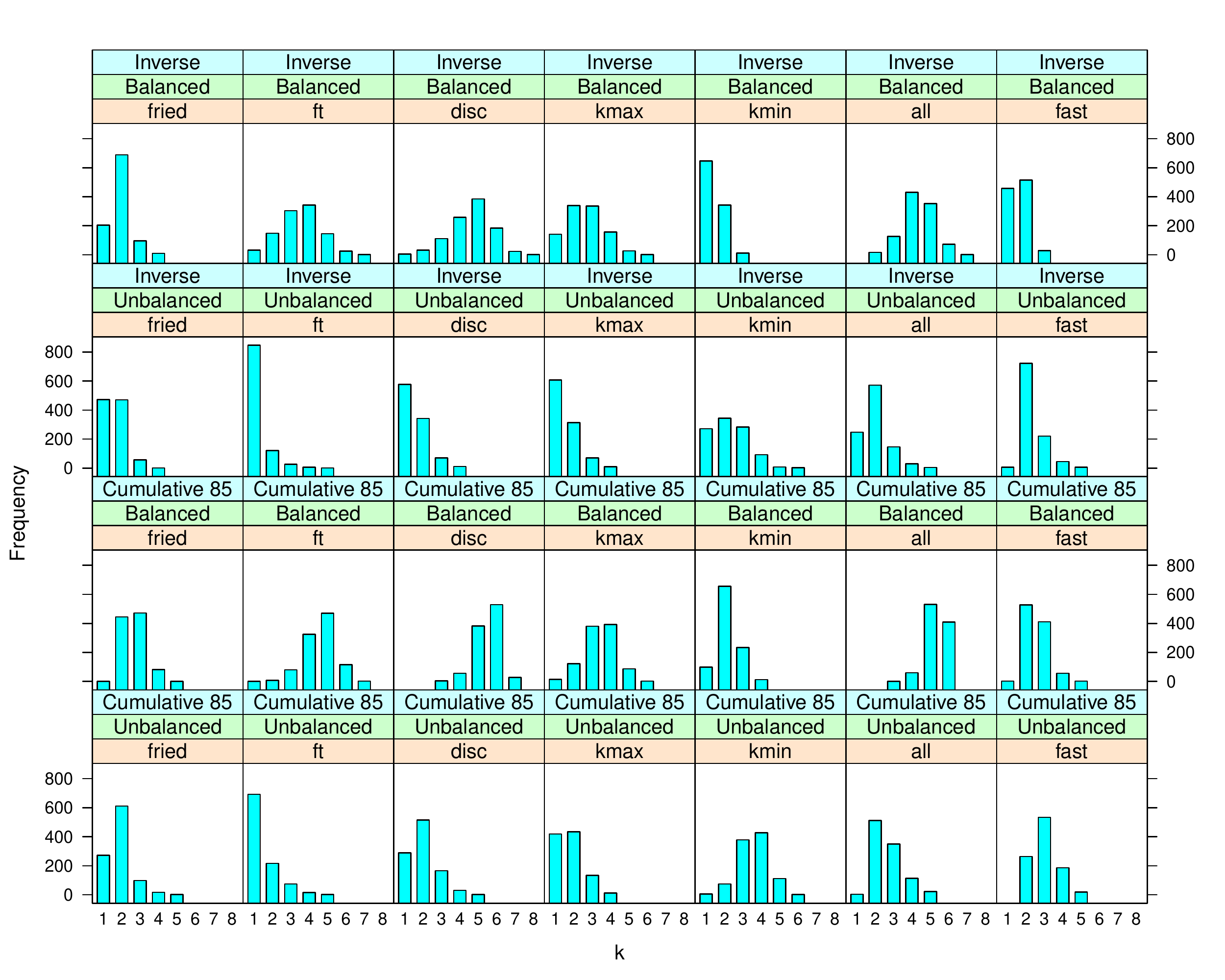}\\
\caption{Number of directions chosen by the different
aggregation methods.} \label{Simuk} \end{figure}

In Figure~\ref{Simuk} is shown how many directions the
different aggregation methods choose. Recall that the number of clusters is 3 and so $k=2$
should be right but naturally from the figure one cannot
conclude if when $k=2$ the right two directions are chosen. The
Figure nevertheless gives the number of directions selected and in general it seems that cum85
chooses more directions than \code{inverse}. 

\begin{figure} \center
\includegraphics[width=1.0\textwidth]{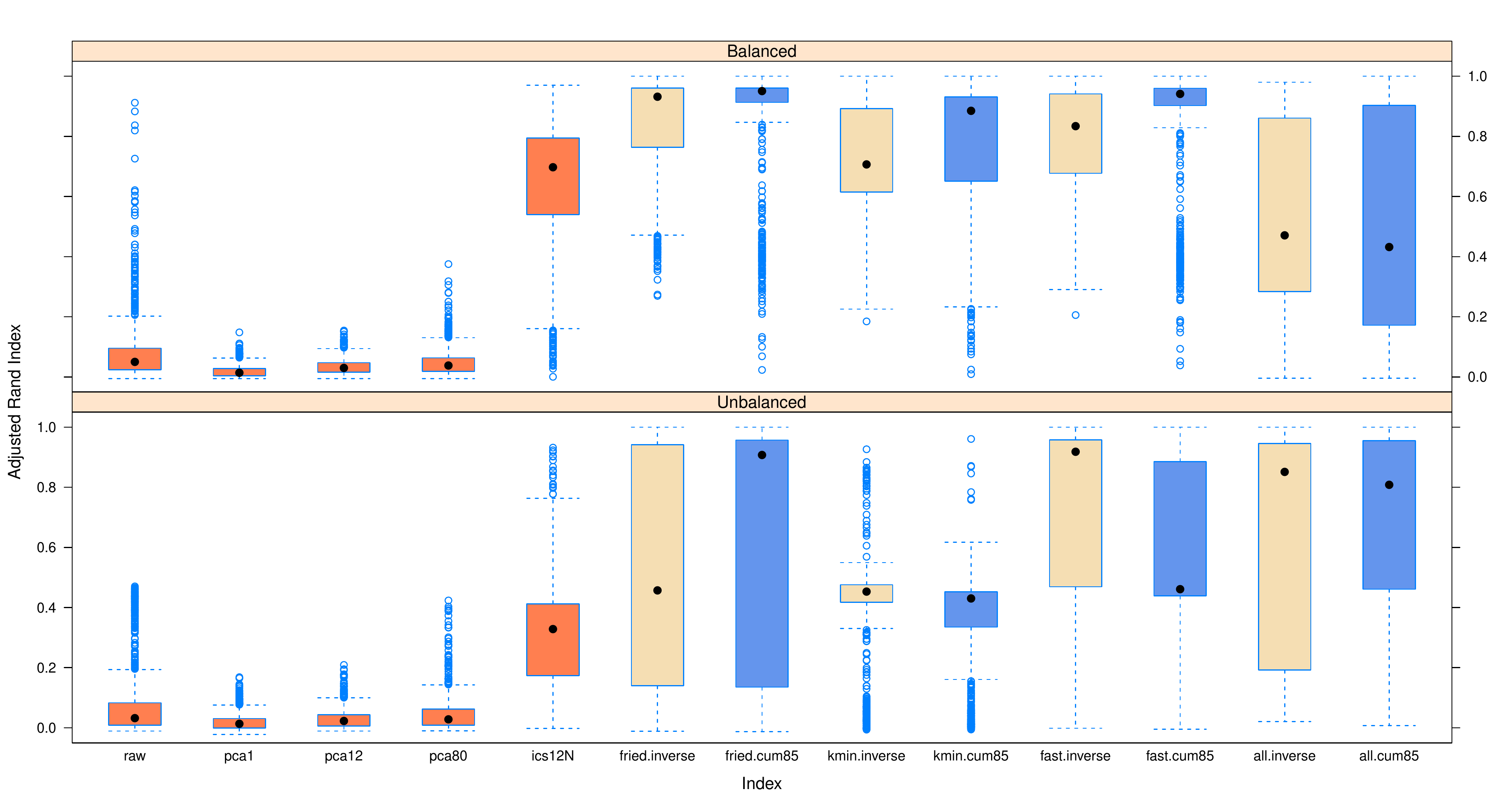}\\
\caption{Adjusted Rand index for k-means underlying different
component selection methods.} \label{SimuComp} \end{figure}

At the end of this Section we put EPP in context to
competing methods. We consider applying k-means to the raw data
using all 10 dimensions as well as principal component
analysis, where we use the first PC, the first two PCs and the first $l$ PCs
which explain 80\% of the variation. We also
compare  EPP to invariant coordinate selection (ICS)
\cite{TylerCritchleyDumbgenOja2009} using the default as
implemented in \pkg{ICS} \cite{NordhausenOjaTyler2008} and
choose the first two and last two components. The corresponding
adjusted Rand index values are compared to our EPP approach in Figure~\ref{SimuComp}.

From this comparison one can  conclude that using PCA in this context is not beneficial at all.
It seems even worse than applying k-means on the raw data. ICS is a clear improvement but still not as good as EPP.
While for the balanced setting the Friedman index is performing best when aggregated using 85\% in the cumulative approach,
for the unbalanced setting the combination of Friedmann and minimizing kurtosis using the inverse aggregation approach seems best.
In general we believe that the best index and aggregating method is data dependent. However combining several projections using the inverse aggregation
seems to be a safe choice.

The whole simulation script as well as all simulation results are available upon request.
The packages used for the simulations were \pkg{REPPlab}, \pkg{MCLUST} \cite{FraleyRafteryEtAl:2012}, \pkg{ICS}  and \pkg{lattice}.

\section[Examples]{Examples}\label{Examples}

The following examples will show how to use the \pkg{REPPlab} package for exploratory data analysis to find clusters and outliers in multivariate data.
All examples are made using the 64-bit version of \proglang{R} 3.3.1 \cite{R323} and the packages \pkg{REPPlab} 0.9.4, \pkg{rJava} 0.9-8 \cite{rJava098},
\pkg{lattice} 0.20-33, \pkg{amap} 0.8-14 \cite{amap} 
and \pkg{LDRTools} 0.2.
Note that \proglang{Java} needs to be installed and that, on \proglang{Windows}, often only the 32-bit version of \proglang{Java} will be installed by default and so the 64-bit version needs to be installed
separately. For reproducibility a random seed is set which will be passed from \proglang{R} to \proglang{Java}, where we are using the 64-bit version of \proglang{Java} 8 Update 66.


\subsection[Detecting clusters]{Detecting clusters}

First we consider how to detect clusters in multivariate data using  the \pkg{REPPlab} package. For this demonstration purpose we first use the
so-called Lubishew data which is available in the package \pkg{amap}. The data set is known to contain three clusters and the goal is to find
all of them assuming that the number of clusters is also unknown. It is a very simple example helpful to understand the different steps of the
strategy that can be implemented using \pkg{REPPlab}. At the end of the subsection, we have a short look at a more complex data set which consists
in 572 olive oil samples from Italy available in the package \pkg{tourr}.\\

We first load the necessary packages and the data and also fix the random seed.


\begin{Schunk}
\begin{Sinput}
> library("REPPlab")
> library("amap")
> set.seed(4567)
> data("lubisch")
> X <- lubisch[, 2:7]
> Class <- lubisch[, 8]
\end{Sinput}
\end{Schunk}

\begin{figure}
\center
   \includegraphics[width=0.6\textwidth]{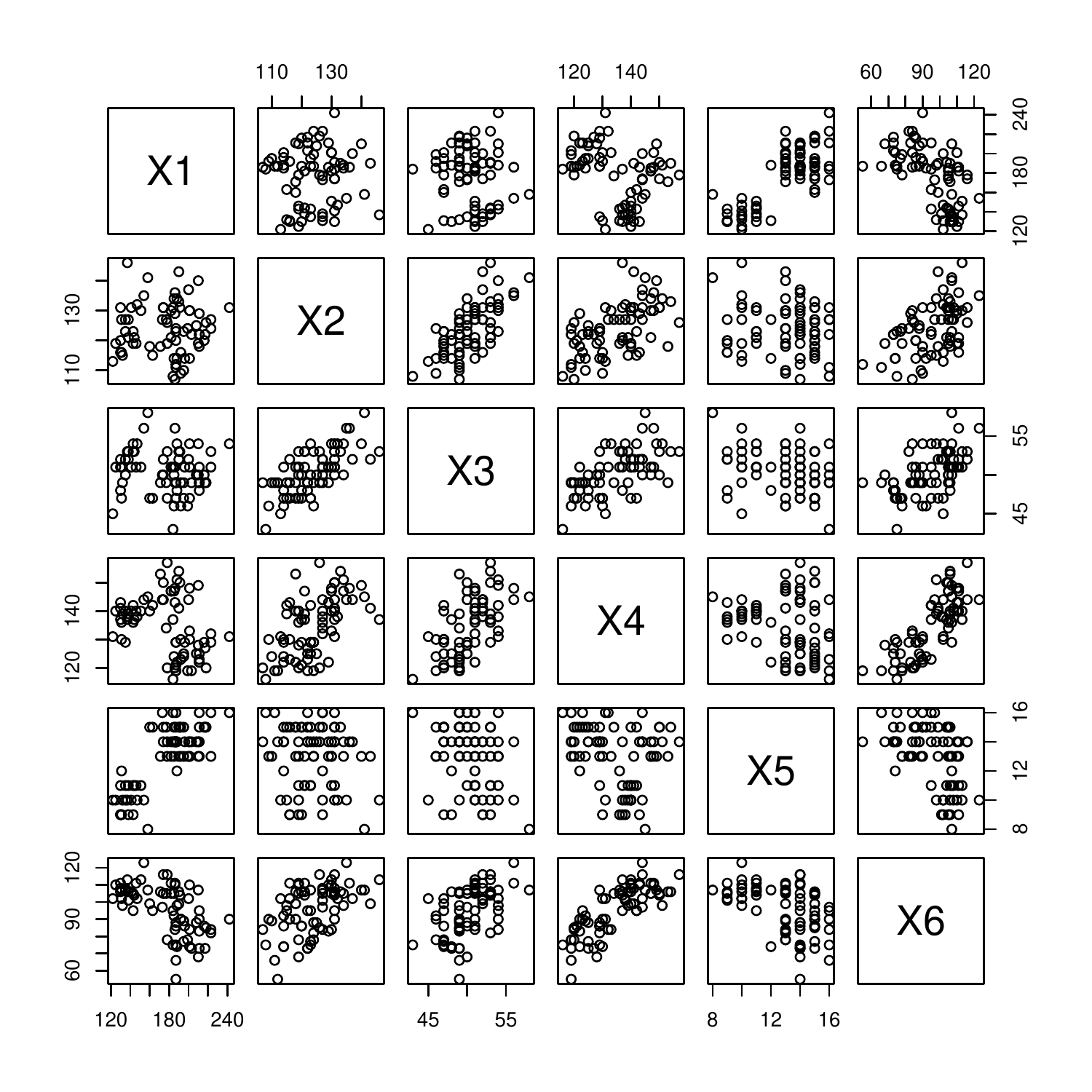}\\
 \caption{Scatterplot matrix of the Lubishew data.} \label{SplotL}
\end{figure}

For convenience we denoted the variables which will be used to identify the clusters as \code{X} and stored the correct class labels in the vector \code{Class}.
Looking at the scatterplot matrix in Figure~\ref{SplotL} does indicate that there are clusters but not so clearly that there are three clusters.

%

\begin{Schunk}
\begin{Sinput}
> pairs(X)
\end{Sinput}
\end{Schunk}

We use now the function \code{EPPlab} to obtain 100 directions using the Friedman index for the sphered data using the Tribes algorithm.


\begin{Schunk}
\begin{Sinput}
> res.Fried.Tribe <- EPPlab(X, PPalg = "Tribe", PPindex =
+       "Friedman", n.simu = 100, maxiter = 200, sphere = TRUE)
> summary(res.Fried.Tribe, digits=4)
\end{Sinput}
\begin{Soutput}
REPPlab Summary
---------------
Index name       : Friedman
Index values     : 0.5517 0.5517 0.5517 0.5517 0.5517 0.5517
0.5517 0.5517 0.5517 0.5517
Algorithm used   : Tribe
Sphered          : TRUE
Iterations       : 48 52 48 53 47 55 42 53 58 46
\end{Soutput}
\end{Schunk}

The \code{summary} is here not too informative but tells us which index and algorithm were used as well as if the data was sphered or not.
It shows also the index value of the first 10 directions and how many iterations they needed. For the first 10 directions the criterion value is always the same, but
to see if different directions are found, plots are more informative.\\

First we look at the screeplot of the \code{res.Fried.Tribe} object

%

\begin{Schunk}
\begin{Sinput}
> screeplot(res.Fried.Tribe, which = 1:100)
\end{Sinput}
\end{Schunk}

\begin{figure}
\center
 \includegraphics[width=0.6\textwidth]{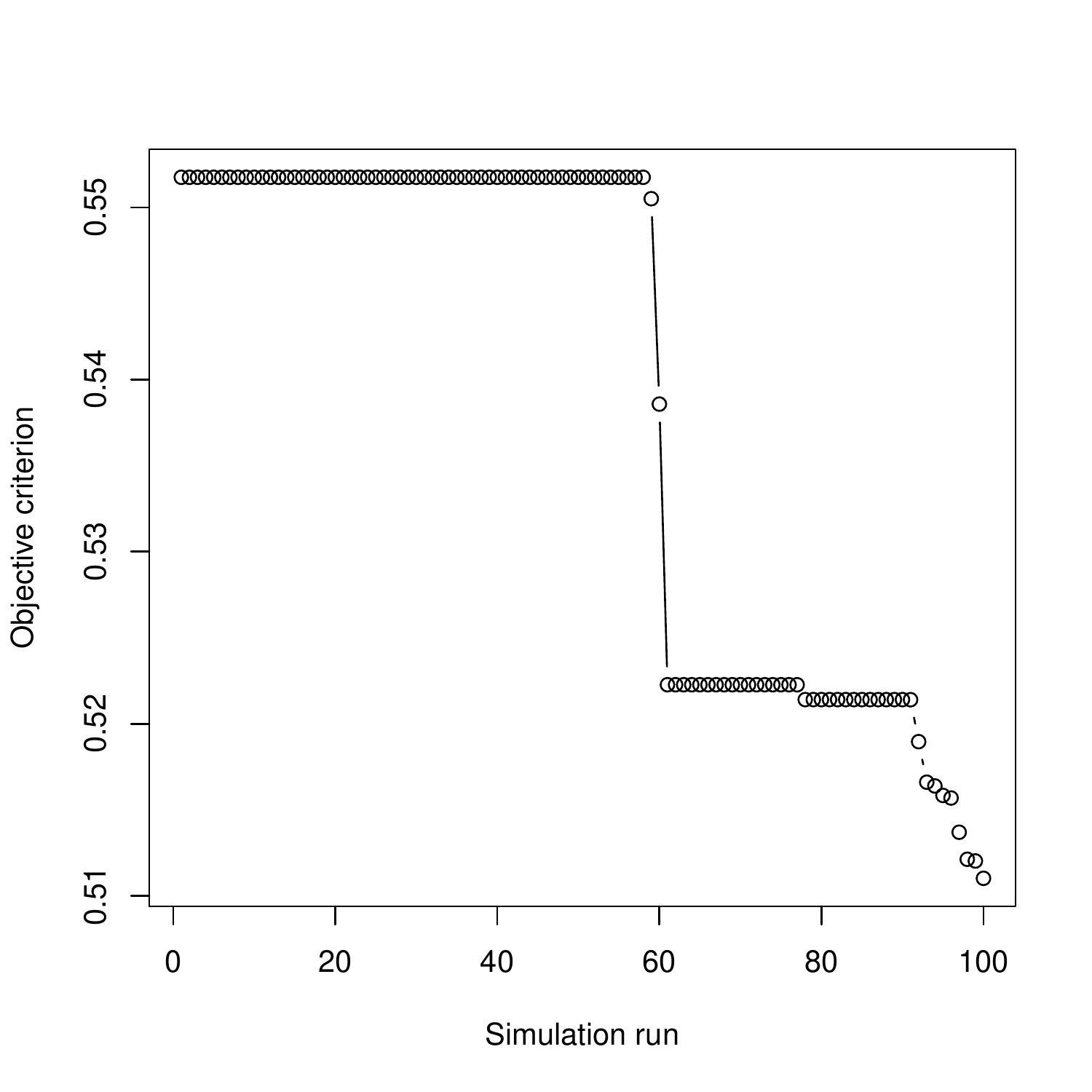}\\
\caption{Screeplot of the criterion values of the Friedman index for the Lubishew data.} \label{ScreeplotL}
\end{figure}

which, as Figure~\ref{ScreeplotL} reveals, gives the same criterion value for around the first 60 directions. Then there is a sharp drop and the next around 20 directions have the
same criterion value before a final small drop occurs for around 10 directions. Finally, the last six directions are associated with some dispersed index values. Naturally,
two identical criterion values do not exclude different directions and the last small drop might also not be a new interesting direction. This can be further investigated by looking at the angles
between the directions.

%

\begin{Schunk}
\begin{Sinput}
> plot(res.Fried.Tribe, type = "angles", which = 1:100)
\end{Sinput}
\end{Schunk}

\begin{figure}
\center
 \includegraphics[width=0.6\textwidth]{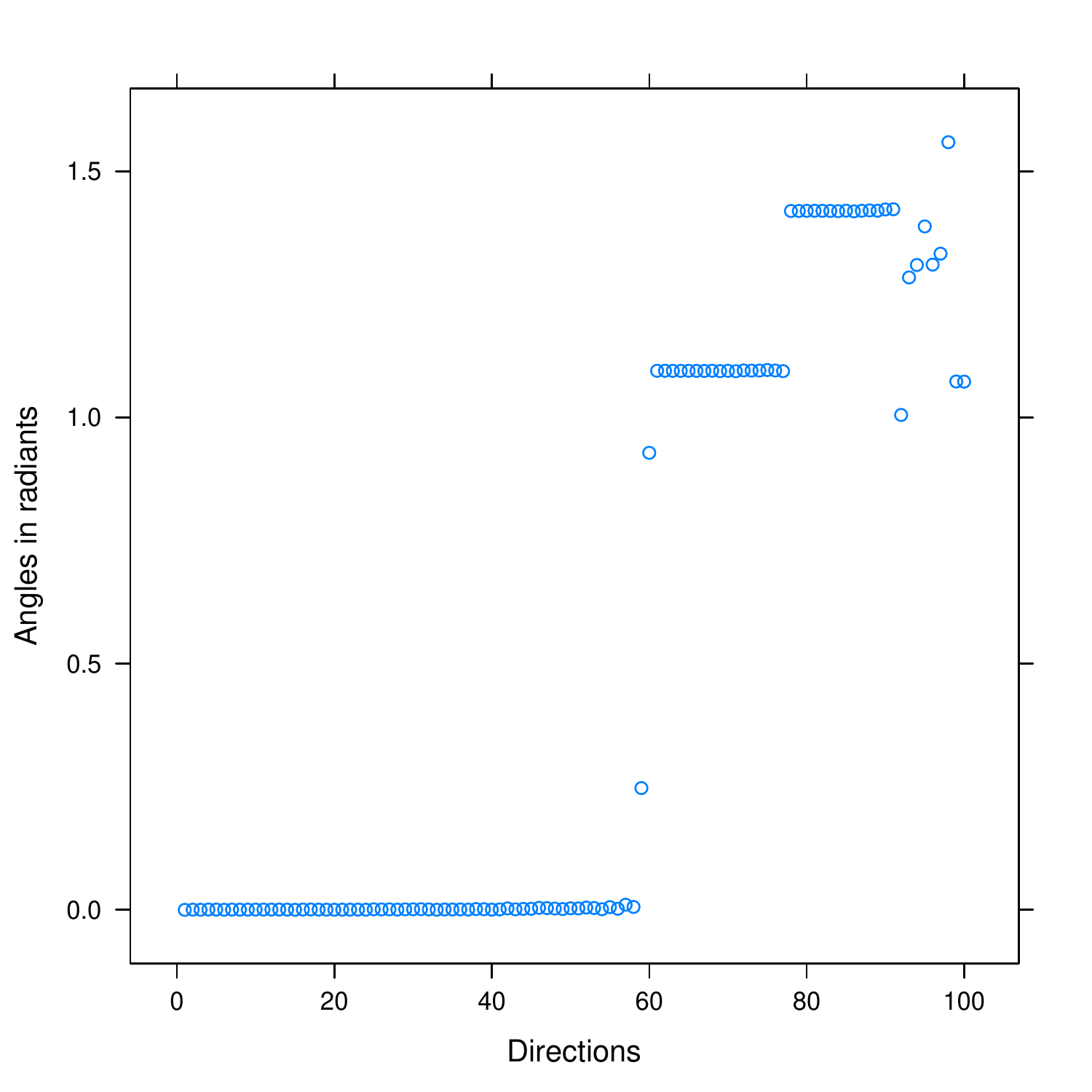}\\
\caption{Angles between the first direction and all other directions for the Lubishew data.} \label{AplotL}
\end{figure}

Figure~\ref{AplotL} shows that the directions in each of the three criterion groups actually are identical and supports the hypothesis that the three different criterion groups correspond to different directions of interest.
This can be verified for example by looking at candidates from all these groups. We choose for this purpose one direction from each of the three groups using Figure~\ref{AplotL}, namely directions 1, 70 and 90 which have the criterion values


\begin{Schunk}
\begin{Sinput}
> res.Fried.Tribe$PPindexVal[c(1, 70, 90)]
\end{Sinput}
\begin{Soutput}
[1] 0.5517 0.5223 0.5214
\end{Soutput}
\end{Schunk}

The code

%

\begin{Schunk}
\begin{Sinput}
> plot(res.Fried.Tribe, which = c(1, 70, 90), layout = c(3, 1))
\end{Sinput}
\end{Schunk}

\begin{figure}
\center
 \includegraphics[width=0.6\textwidth]{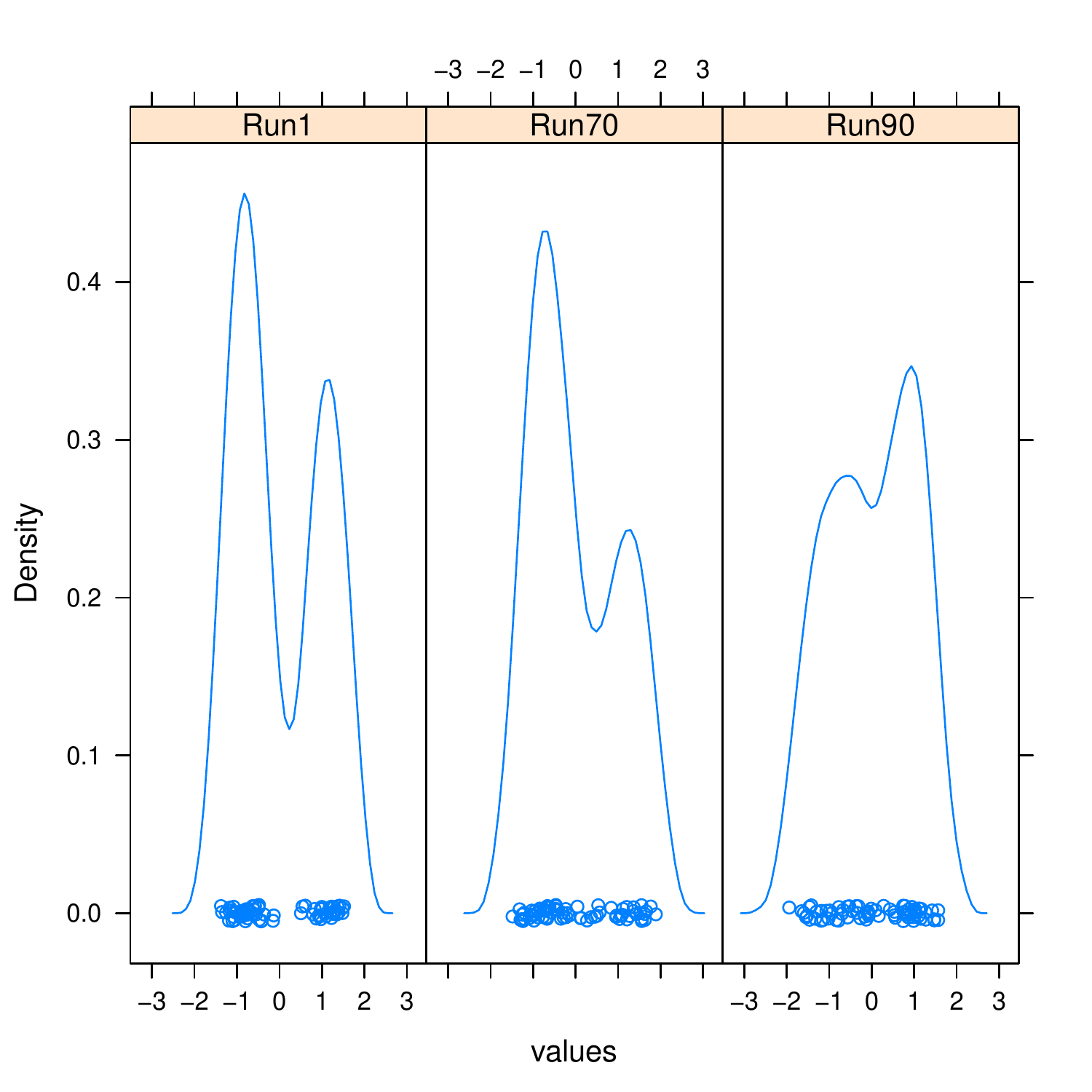}
\caption{Marginal density plots for direction 1, 70 and 90 for the Lubishew data.} \label{DplotL}
\end{figure}

%

\begin{Schunk}
\begin{Sinput}
> pairs(res.Fried.Tribe, which = c(1, 70, 90))
\end{Sinput}
\end{Schunk}

 \begin{figure}
 \center
   \includegraphics[width=0.6\textwidth]{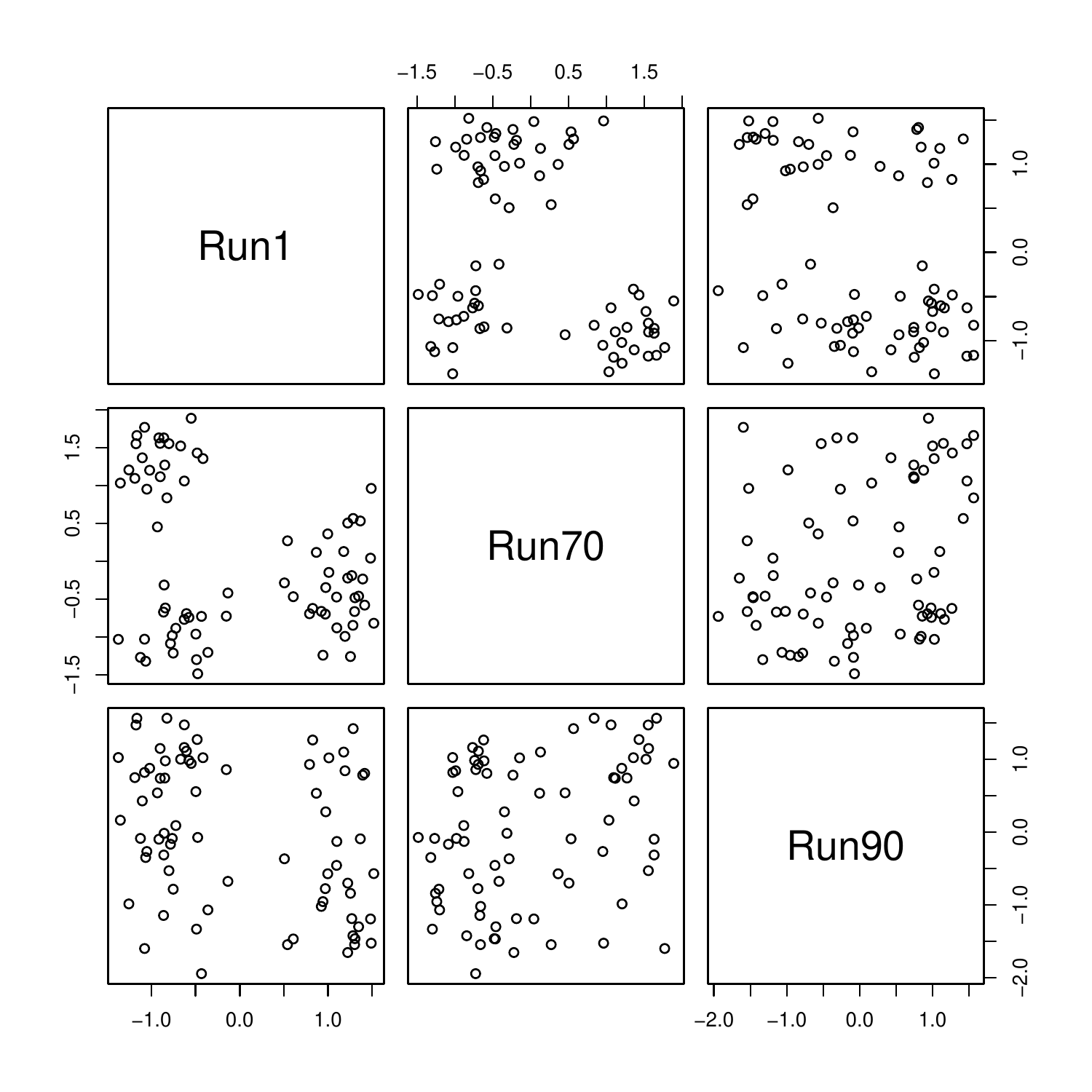}\\
 \caption{Scatter plot matrix of directions 1, 70 and 90 for the Lubishew data.} \label{SCplotL}
 \end{figure}

gives the different marginal kernel density estimates as shown in Figure~\ref{DplotL} and reveals that, at least for direction 1 and 70,
there seem to be two clusters. Looking therefore at the scatter plot matrix of these three directions shows in  Figure~\ref{SCplotL}
that the directions 1 and 70 combined reveal three clusters whereas direction 90 seems not really to add anything of interest.\\

\begin{figure}
\center
   \includegraphics[width=0.6\textwidth]{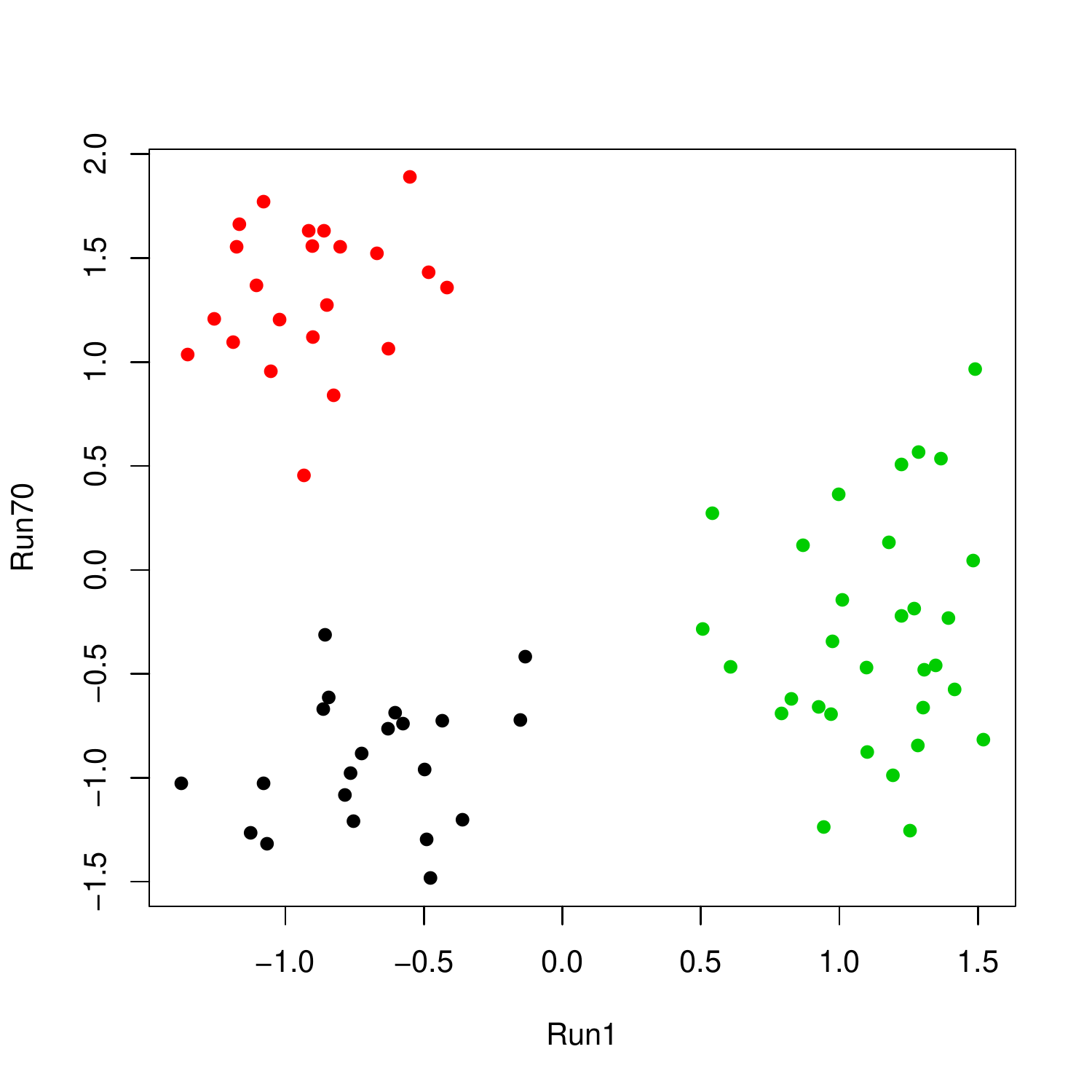}\\
\caption{Scatter plot of directions 1, 70 for the Lubishew data with colored groups.} \label{FplotL2}
\end{figure}

This can be verified as for example shown in Figure~\ref{FplotL2} using the class information labels and the code


\begin{Schunk}
\begin{Sinput}
> plot(fitted(res.Fried.Tribe, which = c(1, 70)), col = Class,
+      pch = 16)
\end{Sinput}
\end{Schunk}

The analysis above is fast to do but focuses on one index only and is difficult to automatize. Following the recently suggested methodology
of \cite{LiskiNordhausenOjaRuizGazen:2015} to summarize many projections, as implemented in the function \code{EPPlabAgg}, provides
therefore a way to combine several indices as already demonstrated in Section~\ref{Simulations}.

For this data, we will then compute, three more indices, each with 100 runs.


\begin{Schunk}
\begin{Sinput}
> set.seed(1234)
> res.Dis.Tribe <- EPPlab(X, PPalg = "Tribe",PPindex =
+   "Discriminant", n.simu = 100, maxiter = 200, sphere = TRUE)
> res.Kmin.Tribe <- EPPlab(X, PPalg = "Tribe", PPindex =
+   "KurtosisMin", n.simu = 100, maxiter = 200, sphere = TRUE)
> res.FT.Tribe <- EPPlab(X, PPalg = "Tribe", PPindex =
+   "FriedmanTukey", n.simu = 100, maxiter = 200, sphere = TRUE)
\end{Sinput}
\end{Schunk}

Note from the previous warning that the convergence criterion was not reached for one simulation run but this has no impact on the final result given the
large number of runs. 




\begin{Schunk}
\begin{Sinput}
> res.ALL <- list(res.Fried.Tribe, res.Dis.Tribe, res.Kmin.Tribe,
+   res.FT.Tribe)
\end{Sinput}
\end{Schunk}

The object \code{res.ALL} is then a combination of all the different indices and too summarize the results we use the function
\code{EPPlabAgg} with the option \code{method = "inverse"} which automatically chooses the rank of the final projection matrices as
recommended in \cite{LiskiNordhausenOjaRuizGazen:2015}.


\begin{Schunk}
\begin{Sinput}
> ALL.agg.inverse <- EPPlabAgg(res.ALL, method = "inverse")
> ALL.agg.inverse
\end{Sinput}
\begin{Soutput}
$P
           [,1]     [,2]      [,3]     [,4]       [,5]
[1,]  0.1348388 -0.10816 -0.175969 -0.27195  0.0004297
[2,] -0.1081595  0.11554  0.005420  0.29396 -0.0580827
[3,] -0.1759687  0.00542  0.952801 -0.05655  0.0013430
[4,] -0.2719504  0.29396 -0.056552  0.78319  0.0225220
[5,]  0.0004297 -0.05808  0.001343  0.02252  0.9960878
[6,] -0.0060522  0.02602 -0.103813  0.07568 -0.0037547
          [,6]
[1,] -0.006052
[2,]  0.026020
[3,] -0.103813
[4,]  0.075682
[5,] -0.003755
[6,]  0.017540

$O
         [,1]     [,2]      [,3]
[1,]  0.13100 -0.03286  0.341465
[2,] -0.09881  0.19532 -0.260050
[3,]  0.06848 -0.75548 -0.614302
[4,] -0.42228  0.51235 -0.585116
[5,] -0.88796 -0.33415  0.309765
[6,] -0.03980  0.12595  0.009663

$k
[1] 3

attr(,"class")
[1] "epplabagg"
\end{Soutput}
\end{Schunk}

%

Hence we can see that, by combining all the 400 one-dimensional projections, three interesting directions are suggested by this method.
Looking at the scatterplot based on these three directions,

\begin{Schunk}
\begin{Sinput}
> pairs(as.matrix(X) %*% ALL.agg.inverse$O)
\end{Sinput}
\end{Schunk}

\begin{figure}
\center
   \includegraphics[width=0.6\textwidth]{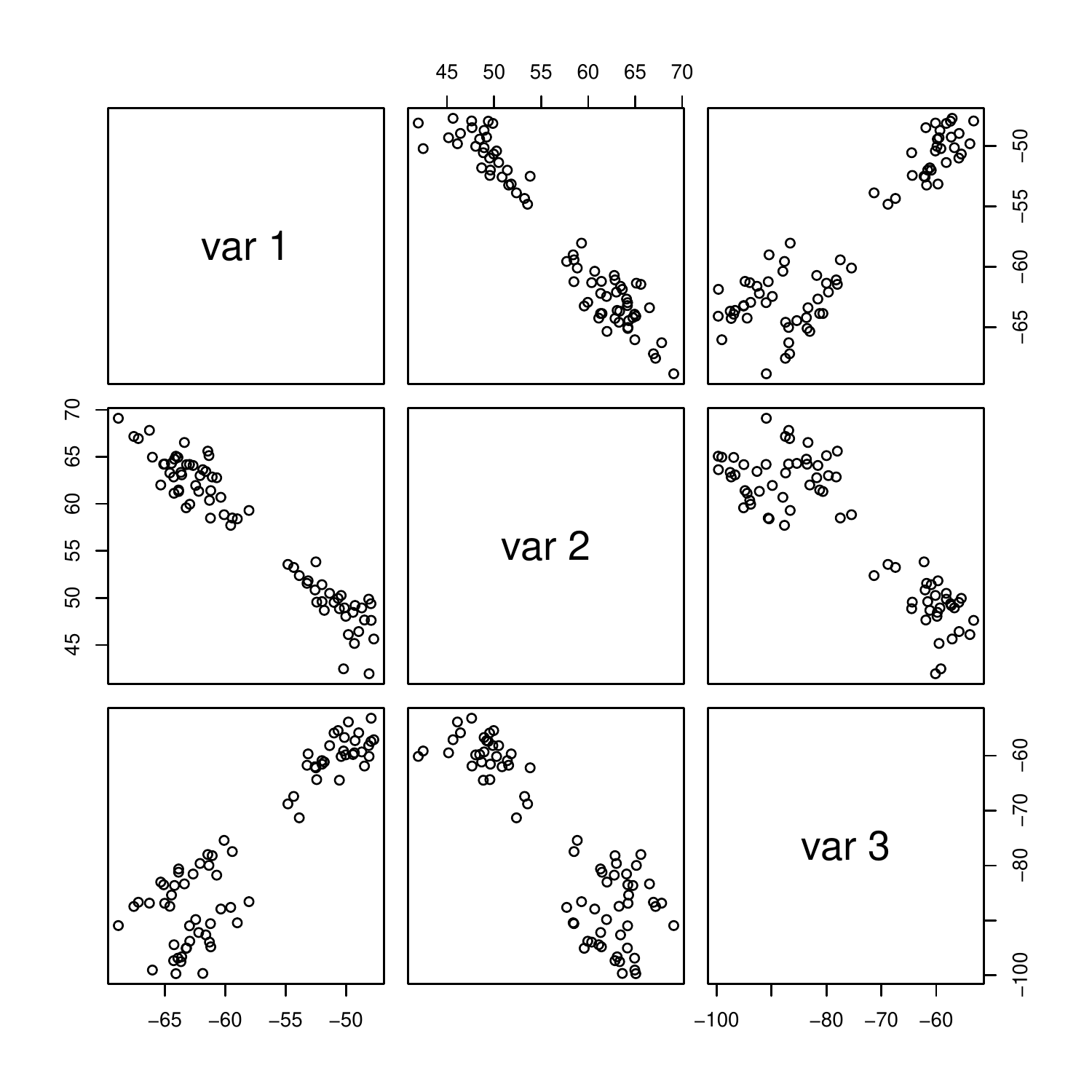}\\
\caption{Scatter plot matrix of three directions derived from \code{AOP} for the Lubishew data.} \label{AOPplot}
\end{figure}

reveals in Figure~\ref{AOPplot} also three clusters. To verify that these are indeed the correct classes we can color them also again
appropriately.

%

\begin{Schunk}
\begin{Sinput}
> pairs( as.matrix(X) %*% ALL.agg.inverse$O, col = Class, pch=16)
\end{Sinput}
\end{Schunk}

\begin{figure}
\center
   \includegraphics[width=0.6\textwidth]{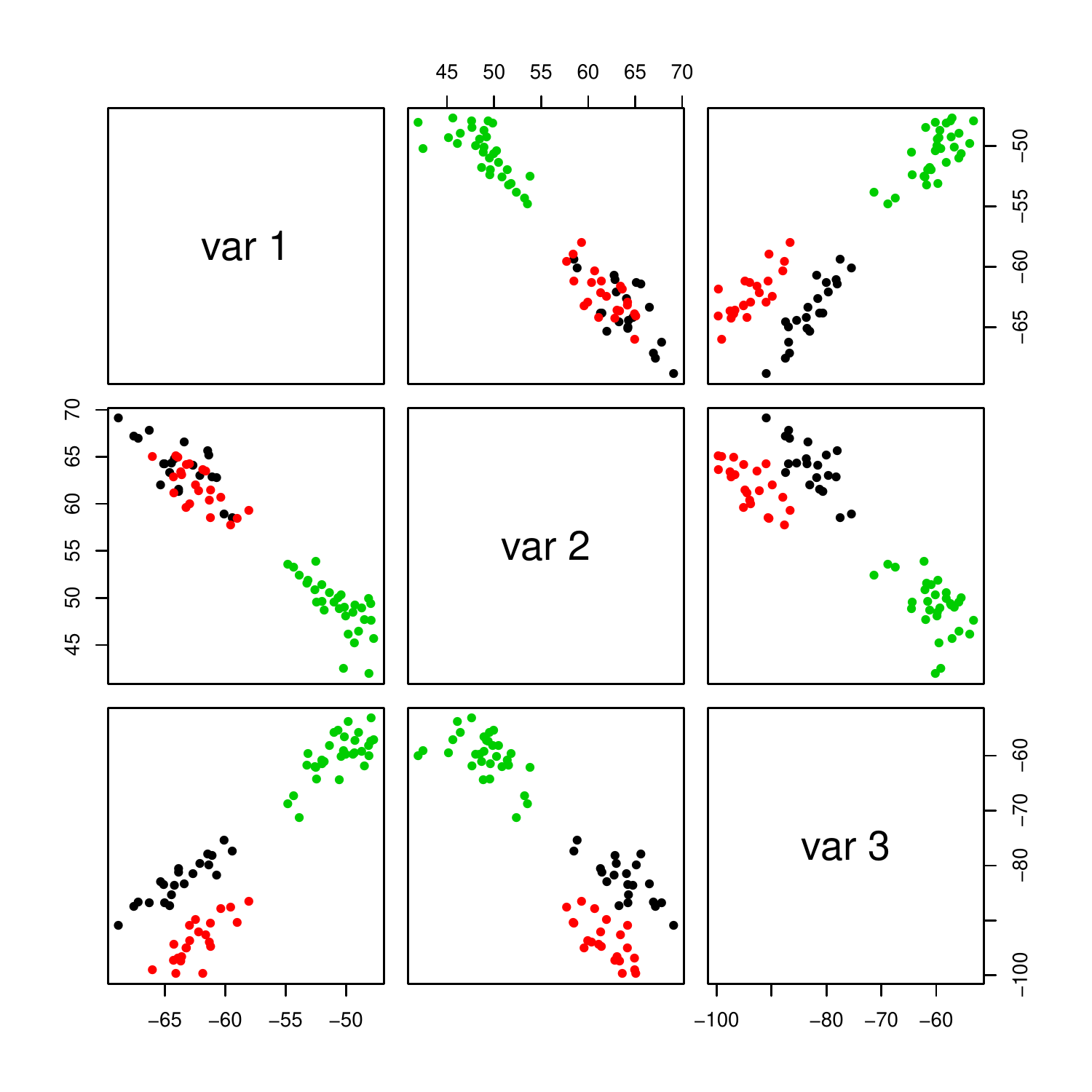}\\
\caption{Scatter plot matrix of three directions derived from \code{AOP} for the Lubishew data, where the different colors correspond to the different classes.} \label{AOPplotcol}
\end{figure}

Figure~\ref{AOPplotcol} reveals that we have found the correct clusters. However, actually for the separation here, the second direction is not useful at all
and also, with the first and second directions, the separation is not as clear as in Figure~\ref{FplotL2} with the two directions 1 and 70.
But the second analysis was more automated and can combine results from different indices. Note that it is also possible to combine projection matrices obtained using different
optimization algorithms.\\

Let us now consider quickly the olive data set which contains 8 fatty acid measurements for 572 olive oil samples. The olive oils are
from different regions from Italy and there are some regional differences in the compositions. The data was
analyzed e.g. in \cite{cook2007} and \cite{berro2010}. For details we refer
to \cite{larabi2016} and just quickly show that applying EPP is here better suited as a preprocessing step for classification compared to for example
PCA.

We load first the data from the \pkg{tourr} and then compute 100 projection directions where we minimize the kurtosis.
\begin{Schunk}
\begin{Sinput}
> library("REPPlab")
> library("tourr")
> data("olive")
> set.seed(1)
> X <- olive[ , 3:10]
> res.Kmin.Tribe <- EPPlab(X, PPalg = "Tribe", PPindex =
+   "KurtosisMin", n.simu = 100, maxiter = 200, sphere = TRUE)
\end{Sinput}
\end{Schunk}

We then look at the corresponding \code{screeplot}
\begin{Schunk}
\begin{Sinput}
> screeplot(res.Kmin.Tribe, which = 1:100)
\end{Sinput}
\end{Schunk}

\begin{figure}
\center
   \includegraphics[width=0.6\textwidth]{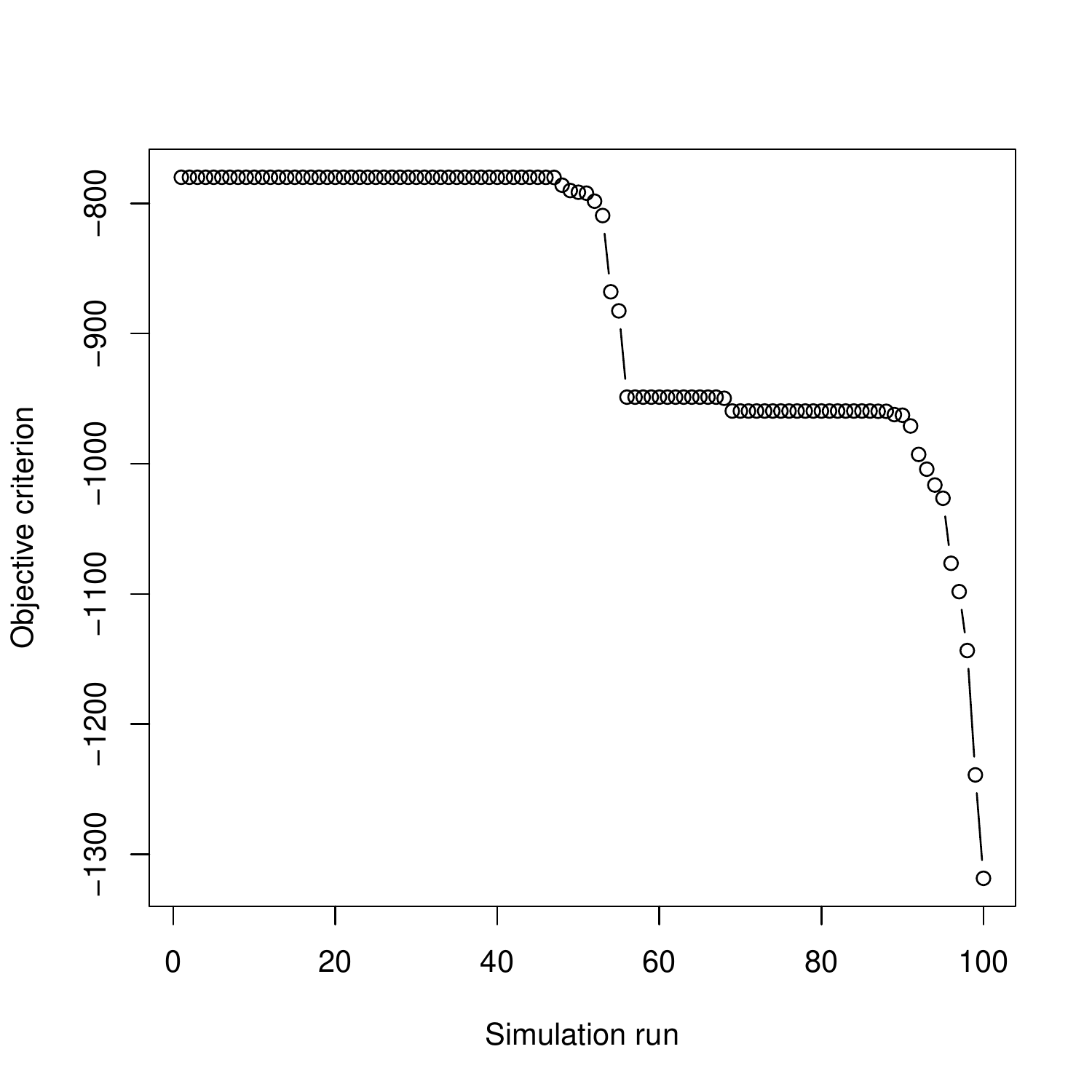}\\
\caption{Screeplot for the \code{olive} data based on 100 directions minimizing kurtosis.} \label{OliveScree}
\end{figure}

given in Figure~\ref{OliveScree}. Based on this figure we choose to look at the kernel densities of directions 1, 60 and 80.

\begin{Schunk}
\begin{Sinput}
>  plot(res.Kmin.Tribe, which = c(1, 60, 80), layout = c(3, 1))
\end{Sinput}
\end{Schunk}

\begin{figure}
\center
   \includegraphics[width=0.6\textwidth]{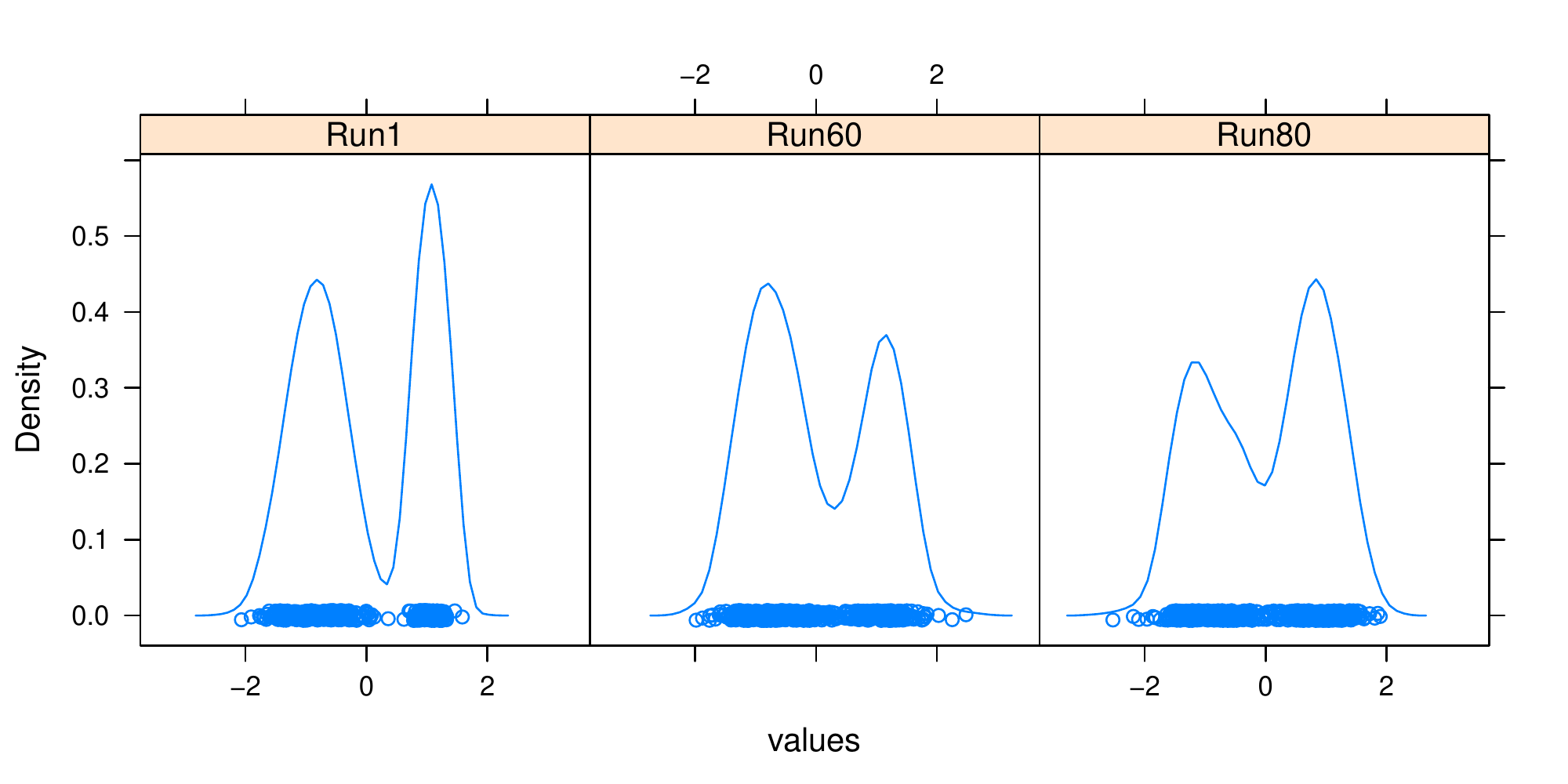}\\
\caption{Kernel densities for directions 1, 60 and 80.} \label{OliveDens}
\end{figure}

All of them show clear clusters and hence we look at the scatterplot matrix of these directions.
\begin{Schunk}
\begin{Sinput}
> pairs(res.Kmin.Tribe, which = c(1, 60, 80))
\end{Sinput}
\end{Schunk}

\begin{figure}
\center
   \includegraphics[width=0.6\textwidth]{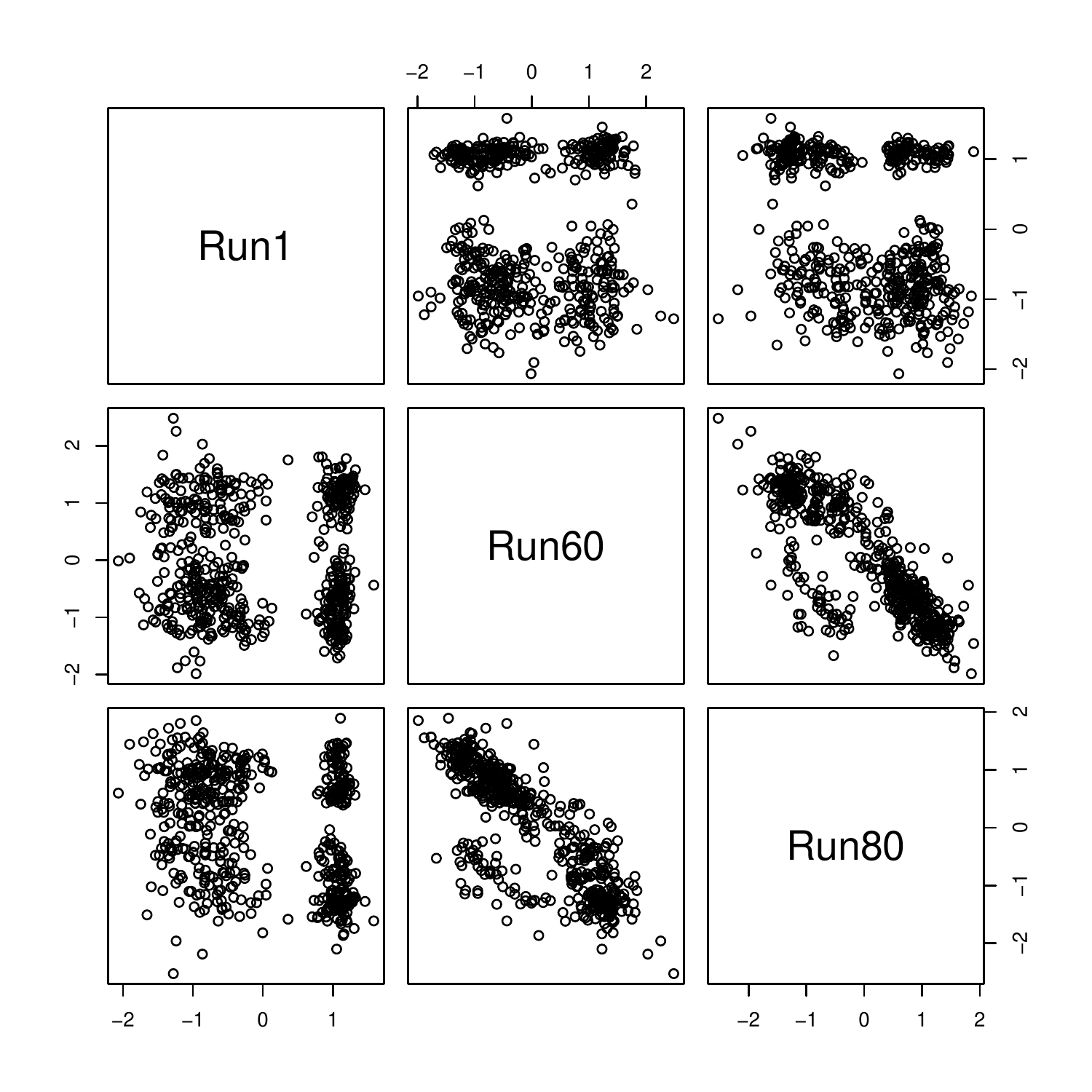}\\
   \caption{Scatter plot matrix of directions 1, 60 and 80.} \label{OliveScat}
\end{figure}

This clearly suggests that there are several clusters in the data. As the reference we show the scatter plot matrix of the first three principal
components, based on the correlation matrix which do no show such clear clusters.

\begin{Schunk}
\begin{Sinput}
> pairs(princomp(X, cor=TRUE)$scores[,1:3])
\end{Sinput}
\end{Schunk}

For a further discussion about applying EPP to this data set see \cite{berro2010} and \cite{larabi2016}.

\begin{figure}
\center
   \includegraphics[width=0.6\textwidth]{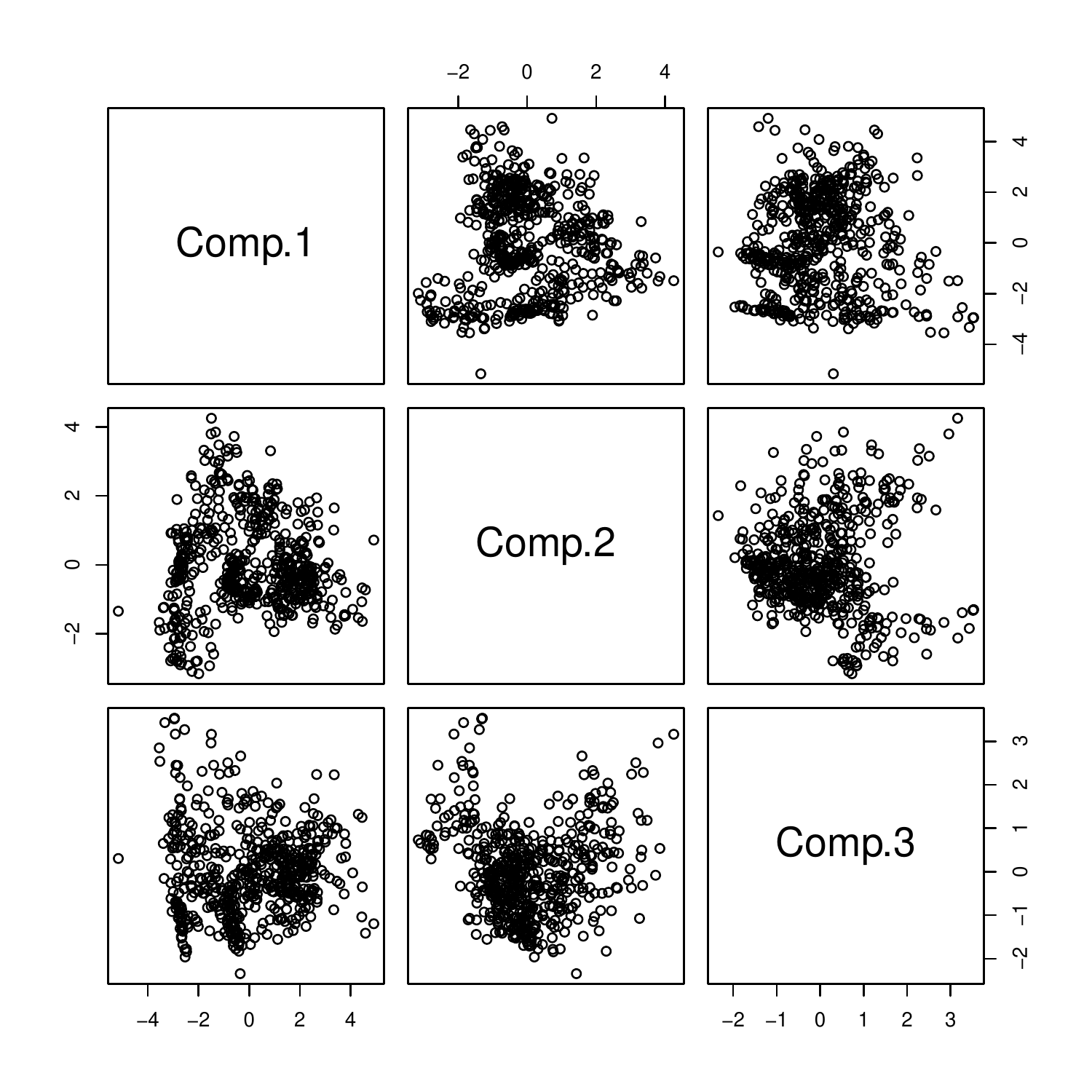}\\
   \caption{Scatter plot matrix of the first three principal components for the \code{olive} data set.} \label{OlivePCA}
\end{figure}

\subsection[Outlier detection example]{Outlier detection example}
In this example the goal is to detect outliers. For the demonstration we use the \code{ReliabilityData} which is made available in \pkg{REPPlab}.
The data provides 55 measurements made on 520 units during one production process. For the producer, it is of interest to find those produced units which might be faulty
in order to check them before selling them.\\

Looking first at all marginal boxplots as shown in
\begin{Schunk}
\begin{Sinput}
> library("REPPlab")
> data("ReliabilityData")
> boxplot(ReliabilityData)
\end{Sinput}
\end{Schunk}

\begin{figure}
\center
   \includegraphics[width=0.6\textwidth]{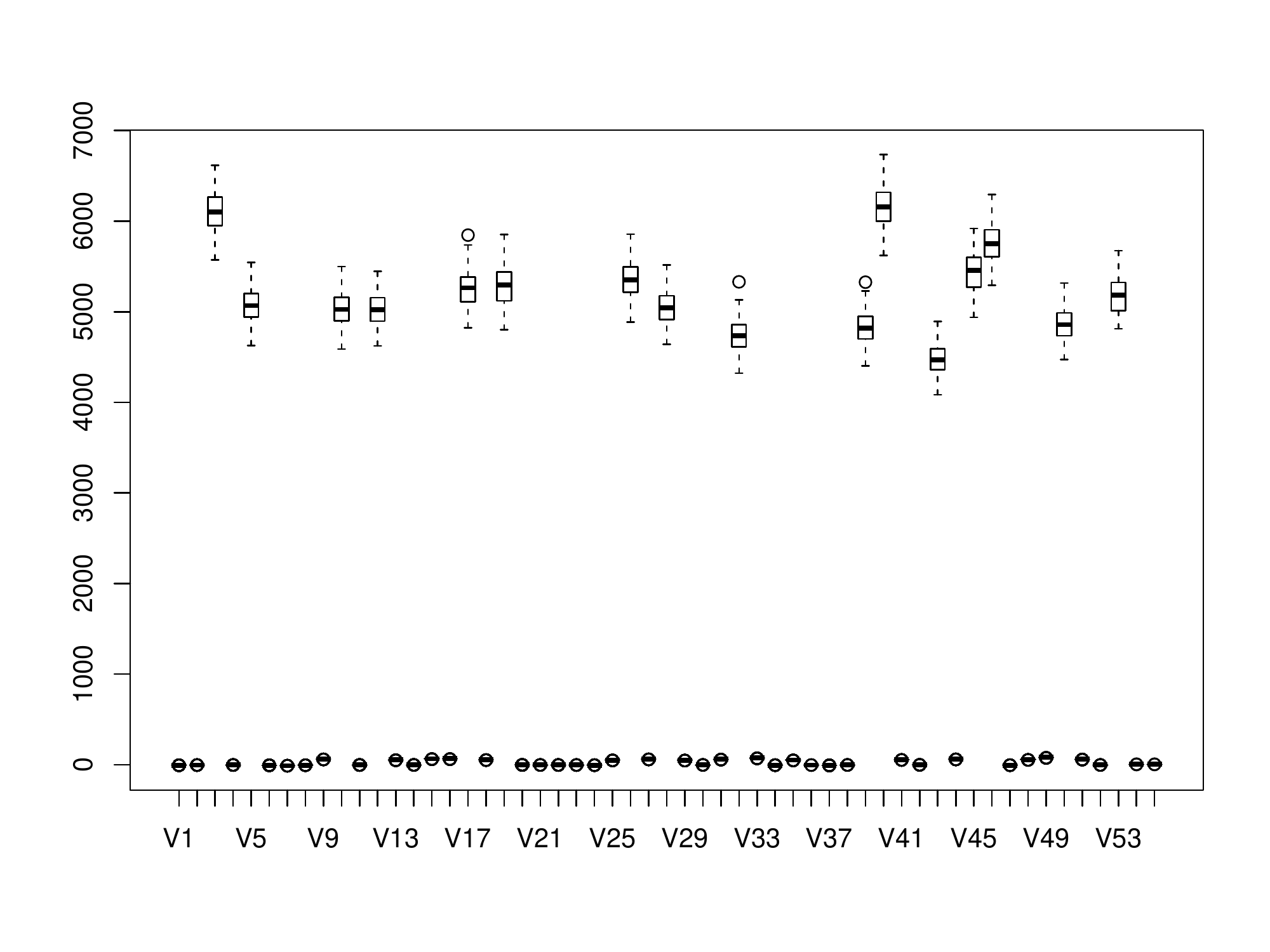}\\
   \caption{Boxplots for all variables in the data set \code{ReliabilityData}.} \label{RelDataBoxPlots}
\end{figure}

reveals, that the scales of the variables differ considerably and shows many marginal outliers.
The exact number of marginal outliers can be obtained for example using the code
\begin{Schunk}
\begin{Sinput}
> id.out <- function(x)
+     { ind <- 1:length(x)
+       BD <- boxplot.stats(x, do.conf = FALSE, do.out = FALSE)
+       ind[x < BD$stats[1] | x > BD$stats[5]]}
> OUT.IDS <- apply(ReliabilityData,2,id.out)
> OUT.IDS.unique <- sort(unique(unlist(OUT.IDS)))
> N.out <- length(OUT.IDS.unique)
> N.out
\end{Sinput}
\end{Schunk}
\begin{Schunk}
\begin{Soutput}
[1] 239
\end{Soutput}
\end{Schunk}

However according to the manufacturer all marginal measurements are in the acceptable range and
 such a huge number of outliers seems unrealistic. Therefore rather multivariate methods should be applied.
For many multivariate methods it is however a problem that several of these variables have almost no variation

\begin{Schunk}
\begin{Sinput}
> round(head(sort(apply(ReliabilityData, 2, sd))), 4)
\end{Sinput}
\begin{Soutput}
    V1    V42    V21    V38    V24    V52
0.0066 0.0095 0.0104 0.0108 0.0121 0.0122
\end{Soutput}
\begin{Sinput}
> round(head(sort(apply(ReliabilityData, 2, mad))), 4)
\end{Sinput}
\begin{Soutput}
   V24    V22     V1    V11    V20    V14
0.0000 0.0033 0.0040 0.0060 0.0068 0.0086
\end{Soutput}
\end{Schunk}
For example methods based on the MCD (Minimum Covariance Determinant), a popular estimate for robust scatter, cannot be computed here for the
whole data set due to these
variables with almost equal measurements. EPP on the other hand does not have such problems as we will demonstrate now.
The recommended projection index for outlier detection is \code{KurtosisMax} for which we will compute 100 runs for this data.


\begin{Schunk}
\begin{Sinput}
> set.seed(4567)
> res.KurtM.Tribe <- EPPlab(ReliabilityData, PPalg = "Tribe",
+   n.simu = 100, maxiter = 200, sphere = TRUE)
\end{Sinput}
\end{Schunk}

The summary of the object \code{res.KurtM.Tribe}
\begin{Schunk}
\begin{Sinput}
> summary(res.KurtM.Tribe)
\end{Sinput}
\begin{Soutput}
REPPlab Summary
---------------
Index name       : KurtosisMax
Index values     : 64529 63631 62662 61471 61455 60359 60260
59984 59885 59371
Algorithm used   : Tribe
Sphered          : TRUE
Iterations       : 68 96 71 109 75 83 72 44 63 76
\end{Soutput}
\end{Schunk}
does not reveal anything useful for this purpose. For outlier detection it is better to call the function \code{EPPlabOutlier}
on the object. To define an outlier in this context we use first the rule that the observation must deviate by more than 5 standard deviations from the mean.

%
%
\begin{Schunk}
\begin{Sinput}
> OUTms <- EPPlabOutlier(res.KurtM.Tribe, k = 5, location =
+   mean, scale = sd)
> summary(OUTms)
\end{Sinput}
\begin{Soutput}
REPPlab Outlier Summary
-----------------------
Index name       : KurtosisMax
Algorithm used   : Tribe
Location used    : mean
Scale used       : sd
k value used     : 5
-----------------------

Number of outliers detected:
 5

Observations considered outliers:
OutlierID:  obs57 obs268 obs414 obs503 obs512
Frequency:  2     1      73     1      79
Percentage: 2     1      73     1      79
\end{Soutput}
\end{Schunk}

\begin{figure}
\center
   \includegraphics[width=0.6\textwidth]{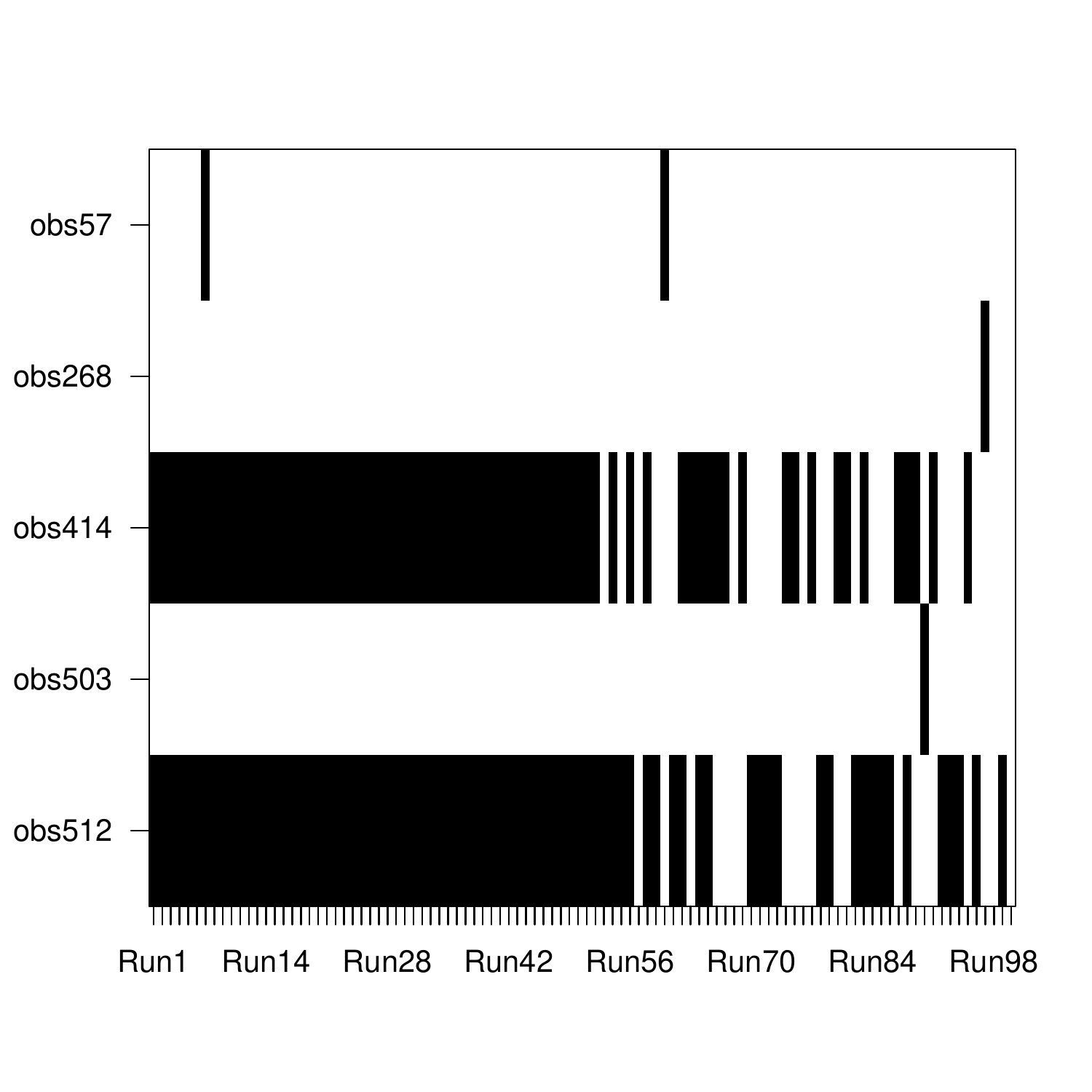}\\
   \caption{Visual representation of the observations considered outliers for the \code{ReliabilityData} when using mean and standard deviation.} \label{OplotMS}
\end{figure}

The summary of this function reveals that five observations are considered outliers in this context. Observations 414 and 512 were extreme in 73 and 79 runs out of the 100 runs
while the other three observations were only rarely considered outliers. As we used 100 runs here, the values for \code{Frequency}
and \code{Percentage} are identical. A graphical display (see Figure~\ref{OplotMS}) of this result is obtained as
%
%

\begin{Schunk}
\begin{Sinput}
> plot(OUTms, las = 1)
\end{Sinput}
\end{Schunk}

In case it is costly to check for production errors, one may require that an observation is an outlier only when it is detected so in at least four directions.


\begin{Schunk}
\begin{Sinput}
> totms <- apply(OUTms$outlier, 1, sum)
> totms[totms > 3]
\end{Sinput}
\begin{Soutput}
obs414 obs512
    73     79
\end{Soutput}
\end{Schunk}

This condition leaves two candidates.

It is however well known that both mean and standard deviation suffer considerably from the presence of outliers. Therefore also robust measures can be used to categorize outliers.
Common choices are for example to replace the mean and the standard deviation with the median and the median absolute deviation respectively.

%
%
%

\begin{Schunk}
\begin{Sinput}
> OUTmm <- EPPlabOutlier(res.KurtM.Tribe, k = 5, location =
+   median, scale = mad)
> summary(OUTmm)
\end{Sinput}
\begin{Soutput}
REPPlab Outlier Summary
-----------------------
Index name       : KurtosisMax
Algorithm used   : Tribe
Location used    : median
Scale used       : mad
k value used     : 5
-----------------------

Number of outliers detected:
 32

Observations considered outliers:
OutlierID:  obs34 obs48 obs57 obs74 obs129 obs135 obs156
Frequency:  2     2     33    2     2      2      1
Percentage: 2     2     33    2     2      2      1

OutlierID:  obs164 obs173 obs174 obs178 obs183 obs196
Frequency:  2      2      2      2      3      2
Percentage: 2      2      2      2      3      2

OutlierID:  obs225 obs259 obs268 obs319 obs326 obs344
Frequency:  1      1      1      3      1      2
Percentage: 1      1      1      3      1      2

OutlierID:  obs355 obs363 obs367 obs372 obs375 obs414
Frequency:  3      1      6      2      1      74
Percentage: 3      1      6      2      1      74

OutlierID:  obs429 obs460 obs461 obs466 obs478 obs503
Frequency:  2      1      2      2      2      1
Percentage: 2      1      2      2      2      1

OutlierID:  obs512
Frequency:  81
Percentage: 81
\end{Soutput}
\end{Schunk}

\begin{figure}
\center
   \includegraphics[width=0.6\textwidth]{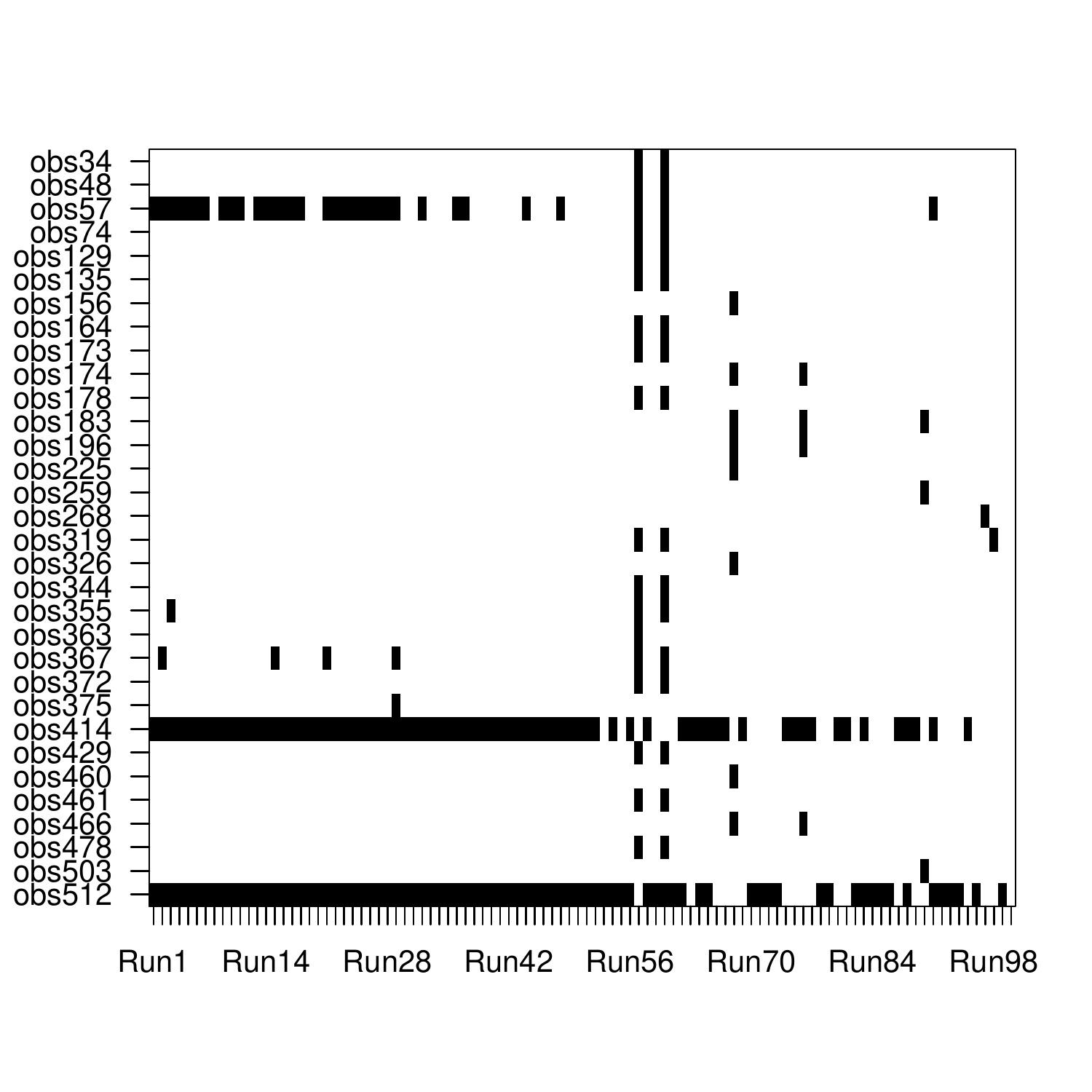}\\
   \caption{Visual representation of the observations considered outliers for the \code{ReliabilityData} when using median and median absolute deviation.} \label{OplotMM}
\end{figure}

In that case, 32 observations are considered outliers. The most extreme ones are, as for the non-robust call, observations 414 and 512.
But also observation 57 is now considered quite often as an outlier. Again we can also visualize this (Figure~\ref{OplotMM}) via

%

\begin{Schunk}
\begin{Sinput}
> plot(OUTmm, las = 1)
\end{Sinput}
\end{Schunk}

Selecting now only those observations which were considered in at least 4 directions as outlier, gives the following observations

\begin{Schunk}
\begin{Sinput}
> totmm <- apply(OUTmm$outlier, 1, sum)
> totmm[totmm > 3]
\end{Sinput}
\begin{Soutput}
 obs57 obs367 obs414 obs512
    33      6     74     81
\end{Soutput}
\end{Schunk}

\begin{figure}
\center
   \includegraphics[width=0.6\textwidth]{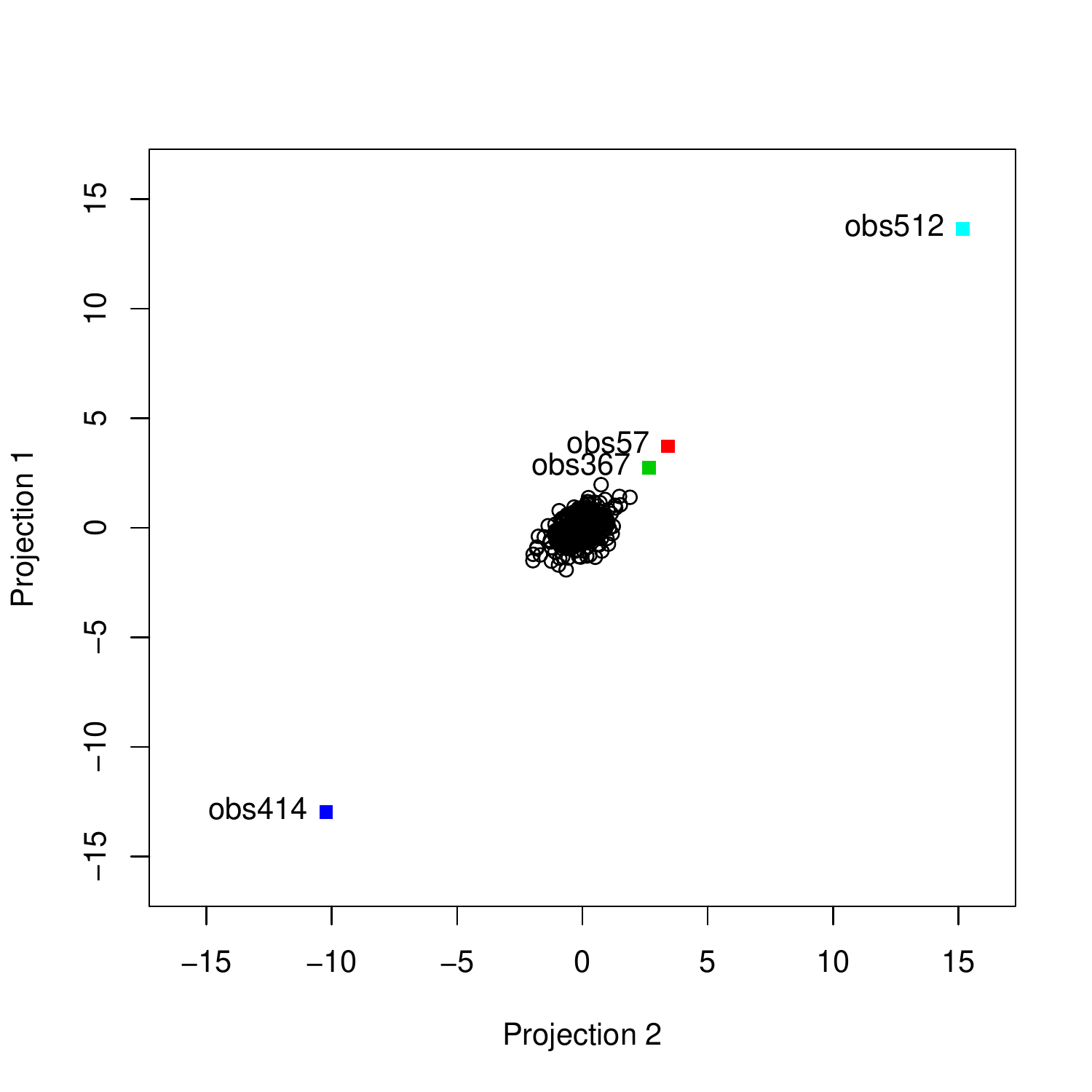}\\
\caption{Visual inspection of the four outlier candidates in the  \code{ReliabilityData} for the first two projections.} \label{Oscatter}
\end{figure}

Hence, next to the same two observations from the previous call, we have to add two more suspects. A scatter plot of the first and second direction (Figure~\ref{Oscatter})
will then be used to see how extreme the observations are in these directions.
%

\begin{Schunk}
\begin{Sinput}
> ProjDir1 <- fitted(res.KurtM.Tribe)
> ProjDir2 <- fitted(res.KurtM.Tribe, which = 2)
> range(ProjDir1)
\end{Sinput}
\begin{Soutput}
[1] -10.22  15.17
\end{Soutput}
\begin{Sinput}
> range(ProjDir2)
\end{Sinput}
\begin{Soutput}
[1] -12.98  13.64
\end{Soutput}
\end{Schunk}

\begin{Schunk}
\begin{Sinput}
> plot(ProjDir1[-c(57, 367, 414, 512)], ProjDir2[-c(57, 367,
+   414, 512)], ylim = c(-16, 16), xlim = c(-16, 16),
+   ylab = "Projection 1",
+   xlab = "Projection 2")
> points(ProjDir1[c(57, 367, 414, 512)], ProjDir2[c(57, 367,
+   414, 512)], col = 2:5, pch = 15)
> text(ProjDir1[c(57, 367, 414, 512)], ProjDir2[c(57, 367,
+   414, 512)], pos = 2, label = row.names(ReliabilityData)
+   [c(57, 367, 414, 512)])
\end{Sinput}
\end{Schunk}

As can be seen in this last figure, the two directions lead to a similar representation and all four observations can be considered as quite extreme with
observations 414 and 512 being more extreme than 57 and 367.

\section[Conclusion]{Conclusion}\label{Conclusion}

EPP is a useful and interesting preliminary step in data analysis that may reveal non-gaussian hidden structures such as clusters or outliers
in multivariate numerical data sets. However, very few EPP tools are available in the standard statistical softwares. The package
\pkg{REPPlab} is a good opportunity for \proglang{R} users to access several projection indices and optimization algorithms that
are already available in the \proglang{Java} program \EPPlab{}. \pkg{REPPlab} also offers some extra functionality to explore and
summarize the obtained
projections. Some of the functionalities, such as the exploration of the projection index values and the cosines between projection
directions and also some outlier detection tools,  were already present in \EPPlab{}. Nevertheless, the novel approach of combining several
projections from many different indices or/and algorithms is original and seems promising in view of the shown simulation results and examples.
The implementation of other PP indices like the Stahel-Donoho index used for instance in \cite{maronna1999} for outlier detection is one of the perspectives of the present work.

\section*{Acknowledgements}
The work of Klaus Nordhausen was supported by the Academy of Finland (grant 268703).
The authors wish to acknowledge CSC -- IT Center for Science, Finland, for providing computational resources.
The article is based upon work from COST Action CRoNoS, supported by COST (European Cooperation in Science and Technology).

\bibliography{REPPlabBib}   

%
%

\end{document}